\documentclass[12pt,preprint]{aastex}

\usepackage{natbib}
\usepackage{graphicx}
\usepackage{wasysym}

\def\lapp{\ifmmode\stackrel{<}{_{\sim}}\else$\stackrel{<}{_{\sim}}$\fi}
\def\gapp{\ifmmode\stackrel{>}{_{\sim}}\else$\stackrel{>}{_{\sim}}$\fi}

\newcommand{\miiia}{PSR~J1342+2822A}
\newcommand{\miiib}{PSR~J1342+2822B}
\newcommand{\miiic}{PSR~J1342+2822C}
\newcommand{\miiid}{PSR~J1342+2822D}
\newcommand{\mva}{PSR~B1516+02A}
\newcommand{\mvb}{PSR~B1516+02B}
\newcommand{\mvc}{PSR~J1518+0204C}
\newcommand{\mvd}{PSR~J1518+0204D}
\newcommand{\mve}{PSR~J1518+0204E}
\newcommand{\mxiiia}{PSR~B1639+36A}
\newcommand{\mxiiib}{PSR~B1639+36B}
\newcommand{\mxiiic}{PSR~J1641+3627C}
\newcommand{\mxiiid}{PSR~J1641+3627D}
\newcommand{\mxiiie}{PSR~J1641+3627E}
\newcommand{\mxva}{PSR~B2127+11A}
\newcommand{\mxvb}{PSR~B2127+11B}
\newcommand{\mxvc}{PSR~B2127+11C}
\newcommand{\mxvd}{PSR~B2127+11D}
\newcommand{\mxve}{PSR~B2127+11E}
\newcommand{\mxvf}{PSR~B2127+11F}
\newcommand{\mxvg}{PSR~B2127+11G}
\newcommand{\mxvh}{PSR~B2127+11H}
\newcommand{\mliiia}{B1310+18}
\newcommand{\mlxxia}{PSR~J1954+1847}
\newcommand{\nvia}{PSR~J1905+0154A}
\newcommand{\nvib}{PSR~J1905+0154B}
\newcommand{\nsixa}{PSR~B1908+00}
\newcommand{\nsixb}{PSR~J1911+0101B}
\newcommand{\terad}{PSR~J1748$-$2446ad}
\newcommand{\msun}{\ifmmode\mbox{M}_{\odot}\else$\mbox{M}_{\odot}$\fi}
\newcommand{\lsun}{\ifmmode\mbox{L}_{\odot}\else$\mbox{L}_{\odot}$\fi}
\newcommand{\rsun}{\ifmmode\mbox{R}_{\odot}\else$\mbox{R}_{\odot}$\fi}
\newcommand{\degrees}{\ifmmode^{\circ}\else$^{\circ}$\fi}
\newcommand{\amin}{\ifmmode^{\prime}\else$^{\prime}$\fi}
\newcommand{\asec}{\ifmmode^{\prime\prime}\else$^{\prime\prime}$\fi}

\shorttitle{MSPs in GCs with Arecibo}
\shortauthors{Hessels et al.}

\begin{document}

\title{A 1.4-GHz Arecibo Survey for Pulsars in Globular Clusters}

\author{J.~W.~T. Hessels$^{1,*}$, S.~M. Ransom$^2$, 
I.~H. Stairs$^3$, V.~M. Kaspi$^1$, and P.~C.~C. Freire$^4$}
\affil{$^1$Department of Physics, McGill University, Montreal, QC
  H3A 2T8, Canada; hessels@physics.mcgill.ca}
\affil{$^2$National Radio Astronomy Observatory, 520 Edgemont Road,
Charlottesville, VA 22903}
\affil{$^3$Department of Physics and Astronomy, University of British
Columbia, 6224 Agricultural Road, Vancouver, BC V6T 1Z1, Canada}
\affil{$^4$NAIC, Arecibo Observatory, HC03, Box 53995, Arecibo, PR 00612}
\affil{$^*$Current Address: Astronomical Institute ``Anton Pannekoek'',
University of Amsterdam, Kruislaan 403, 1098 SJ Amsterdam, The Netherlands}

\begin{abstract}
We have surveyed all 22 known Galactic globular clusters observable with the
Arecibo radio telescope and within 70\,kpc of the Sun for radio pulsations
at $\sim 1.4$\,GHz. Data were taken with the Wideband Arecibo Pulsar
Processor, which provided the large bandwidth and high time and frequency
resolution needed to detect fast-spinning, faint pulsars.  We have also
employed advanced search techniques to maintain sensitivity to short orbital
period binaries. These searches have discovered 11 new millisecond pulsars
and 2 promising candidates in 5 clusters, almost doubling the population of
pulsars in the Arecibo-visible globular clusters. Ten of these new pulsars
are in binary systems, and 3 are eclipsing.  This survey has discovered
significantly more very fast-spinning pulsars ($P_{\rm spin} \lesssim
4$\,ms) and short orbital period systems ($P_{\rm orb} \lesssim 6$\,hr) than
previous surveys of the same clusters.  We discuss some properties of these
systems, as well as some characteristics of the globular cluster pulsar
population in general, particularly its luminosity distribution.
\end{abstract}


\section{Introduction}

Currently, there are approximately 130 known pulsars in globular clusters
(GCs)\footnote{A catalog of GC pulsars is maintained by P.~C.~C. Freire at
http://www.naic.edu/$\sim$pfreire/GCpsr.html.}, of which about 60\% are
observed to be in binary systems.\footnote{Note however that, because binary
pulsars are more difficult to detect than isolated pulsars, the observed
binary fraction is a lower limit on the intrinsic binary fraction of the
population.} Roughly two thirds of all the pulsars known in GCs have been
discovered in only the last seven years by surveys
\citep[e.g.][]{pdm+03,rhs+05b,hrs+06} using low-temperature receivers
($T_{\rm rec} \lapp 35$\,K) at central observing frequencies between
$\nu_{\rm center} = 1-$2\,GHz, large bandwidth and high time and
frequency-resolution backends (e.g. the Wideband Arecibo Pulsar Processor,
Dowd, Sisk, \& Hagen 2000\nocite{dsh00}, and the GBT Pulsar Spigot, Kaplan
et al. 2005\nocite{kel+05}), advanced search techniques for binaries
\citep*[e.g.][]{rem02,rce03,cha03}, and copious amounts of processing time on
dedicated computer clusters.  Though a few non-recycled pulsars have been
found in GCs \citep[e.g. PSR~B1718$-$19 in NGC~6342,][]{lbhb93}, almost all
known GC pulsars are MSPs. In fact, the 150 known GCs\footnote{An online 
catalog of Milky Way GCs \citep{har96} is maintained at
http://physun.physics.mcmaster.ca/Globular.html.} orbiting the Milky Way
contain roughly three orders of magnitude more observed millisecond pulsars
(MSPs) per unit mass than the Galactic plane, which contains approximately
60 known MSPs.  GCs have proven to be the most fruitful place to look for
MSPs partly because of the enormous stellar densities in their cores, which
exceed those in the Galactic plane by up to six orders of magnitude. These
conditions promote many different formation processes that create binary
systems in which a neutron star can be spun-up, or ``recycled''
\citep{acrs82,rs82}, through the accretion of matter from its companion star
\citep[see][the most up-to-date general overview of pulsars in GCs, for a
review of the formation and evolutionary processes at work in the cores of
GCs]{cr05}.  Furthermore, the cores of GCs, where most of the MSPs reside,
typically have radii less than an arcminute, small enough to be covered by a
single telescope pointing. This affords the possibility of making single,
deep multi-hour integrations for multiple faint MSPs in GCs, something that
is not feasible in large area surveys of the field.

Some clusters contain many pulsars: Terzan~5 and 47~Tucanae harbor 33 and 22
known pulsars, respectively.  Together, they contain $43$\% of the
total known GC pulsar population \citep{clf+00,rhs+05b,hrs+06}. Finding
numerous pulsars in a single cluster allows interesting studies of the
cluster itself, in addition to the individual pulsars contained therein.
Such studies have included the detection of intra-cluster ionized gas
\citep{fkl+01}, high mass-to-light ratios \citep{dpf+02}, and hints at the
cluster's dynamical history \citep{pdm+03}. However, many clusters still
contain no known pulsars at all, despite sensitive searches \citep{cr05}. In
some cases, this may simply be because pulsars are generally intrinsically
weak objects and GCs are often distant (approximately 90\% of the GCs
orbiting the Milky Way are $> 5$\,kpc from the Sun). Interstellar
scattering, which broadens pulsations because of multi-path propagation, can
also be a major obstacle, especially for MSPs in clusters at low Galactic
latitudes.  At low radio frequencies ($\lesssim 500$\,MHz), scattering can
completely wash out the signal of fast-spinning pulsars. 

While GCs are clearly the most profitable targets for finding MSPs, these
searches remain non-trivial.  First, the requisite short sampling times ($<
100$\,$\mu$s), high frequency resolution ($< 0.5$\,MHz channels), long
integrations (few hours), and large bandwidth ($> 100$\,MHz) of these data
can make data acquisition and storage formidable (data rates $\sim
30-$100\,GB/hr).  Recent surveys with relatively good sensitivity to binary
pulsars have revealed that the majority of GC MSPs are in binaries. The
pulsed signal from these binary pulsars is smeared in Fourier space by their
orbital motion, and thus even a bright pulsar can go undetected if nothing
is done to correct for this modulation over the course of a long
observation. Advanced and computationally intensive techniques, which
partially search orbital parameter space, are required to recover the
majority of the lost signal. The extra effort required to find binaries is
well justified however, as some of the most exotic pulsar binaries known
have been found in GCs.  For instance, PSR~B1620$-$26 in the cluster M4 is
in a hierarchical triple system with a white dwarf and a 1$-$3\,${\rm
M}_{\rm Jup}$ planet, the only planet known in a GC \citep{tacl99,st05};
PSR~J0514$-$4002 in NGC~1851, with an eccentricity of 0.89, is one of the
most eccentric binary pulsars known \citep{fgri04}; PSR~B2127+11C in M15 is
a rare double neutron star binary \citep{agk+90}; \terad\ in Terzan~5 is the
fastest-spinning neutron star known \citep{hrs+06}; and a few MSPs with
possible main-sequence companions have been found \citep[e.g.
PSR~J1740$-$5340,][similar systems have not been found in the Galactic
plane]{dpm+01}.  Other exotic binaries, perhaps even an MSP-MSP binary or a
MSP-black-hole binary, may have effective formation channels only in GCs
\citep*[but see][]{spn04}.

Here we present searches for radio pulsations from 22 GCs, using Arecibo at
central observing frequencies between 1.1$-$1.6\,GHz.  Roughly half of these
clusters have been searched previously with Arecibo at 430\,MHz
\citep{and93,wak+89a,wkm+89b}. The highly successful \citet{and93} survey
found 11 of the 15 pulsars known in these clusters prior to the survey
presented here.  Our searches have uncovered 11 new MSPs and 2 promising
candidates in five clusters: M3, M5, M13, M71, and NGC~6749. Acceleration
searches were crucial in finding all but 2 of these systems.  Ten of the new
pulsars are in binaries, 3 of which are eclipsing, with orbital periods of
only a few hours.  For comparison, no eclipsing systems were known in these
clusters prior to our survey and only 5 of the 15 previously known pulsars
were in binaries.  Of the pulsars presented here, 9 out of 11 have $P_{\rm
spin} < 4$\,ms.  Only 2 of the 15 previously known pulsars in these clusters
have $P_{\rm spin} < 4$\,ms, clearly demonstrating the improved sensitivity
of this survey over past surveys to the fastest MSPs.   In \S2 we describe the
targets, observational setup, and sensitivity of the survey. In \S3 we
outline the search procedure and analysis pipeline. In \S4 we present the
results of the survey.  In the discussion of \S5, we comment on the
characteristics of the GC pulsar population in general, particularly its
luminosity distribution.  In \S6, we conclude.

\section{Observations}

\subsection{Targets}

We observed every known Galactic GC visible from Arecibo\footnote{The
declination range visible from Arecibo is approximately $-1^{\circ}$ to
$+38^{\circ}$.} and within 70\,kpc of the Sun without any selection bias
towards larger or denser clusters. The sample of 22 GCs is listed in
Table~\ref{gcs.tab}, along with basic and derived cluster parameters
\citep[][unless otherwise indicated, all GC quantities used in this paper
are from the 2003 February revision of the catalog]{har96}.  The numbers of
known isolated and binary pulsars in each cluster are indicated, with
figures in parentheses denoting the number of pulsars found by this survey.
For clusters containing known pulsars, the average dispersion measure (DM)
of the pulsars is indicated, as well as their spread in DM, which is in
parentheses (for clusters where two or more pulsars have been found). For
clusters with no known pulsar, the DM is also unknown; the values listed are
derived from the \citet{cl02spec} ``NE2001'' model for the distribution of
free electrons in the Galaxy, using the \citet{har96} position and distance
to the cluster.  

\subsection{Data Acquisition}

Each cluster was observed at least twice for the full time it is visible
with Arecibo (Table~\ref{gcs.tab}).  The clusters were observed in one of
two campaigns in the summers of 2001 and 2002 using the Gregorian L-band
Wide receiver\footnote{For the initial searches, we used the original
``L-Wide'' receiver installed after the Arecibo upgrade in the 1990s.  This
is not the L-band Wide receiver currently available at Arecibo, which was
installed in 2003 February and was used for most of our timing observations.
Though having design differences, these two receivers have comparable
sensitivity.} ($T_{\rm rec} \sim 35$\,K). Depending on the known or
predicted DM of the cluster, the central observing frequency was either
1175\,MHz (DM $\lesssim 100$\,pc cm$^{-3}$) or 1475\,MHz (DM $\gtrsim
100$\,pc cm$^{-3}$). Our observations were made using the Wideband Arecibo
Pulsar Processor \citep[WAPP, see][for details]{dsh00}, a digital
auto-correlator with configurable sampling time (3 or 9-level samples) and
number of lags. Generally, 3-level samples were autocorrelated with 256
lags, accumulated every 64-$\mu$s, and summed in polarization before being
written to the WAPP disk array as 16-bit numbers.  For the few clusters with
known or predicted DMs greater than 100\,pc cm$^{-3}$, we used 128-$\mu$s
sampling and 512 lags.  These configurations were chosen in order to
minimize dispersive pulse smearing and to take full advantage of the WAPP's
maximum sustainable data rate at the time of the observations (8\,MB/s, or
$\sim 30$\,GB/hr). Data were transferred to DLT magnetic tape for offline
analysis and archiving.  These observations resulted in about 4\,TB of data
on about 100 tapes.

At the time of our original cluster search observations, only one WAPP
backend was available, providing 100\,MHz of bandwidth. In more recent
observations of M3, M5, M13, M71, and NGC~6749 (after December 2002), which
were made as part of the timing observations of the new discoveries, we used
three of the four available WAPP backends, each with 100\,MHz of bandwidth
centered at 1170, 1420 and 1520\,MHz.  The frequency gap between the lower
and upper bands was to avoid persistent and intense radio frequency
interference (RFI) in the frequency ranges 1220$-$1360\,MHz and $>
1570$\,MHz.  Using the PRESTO\footnote{See
http://www.cv.nrao.edu/$\sim$sransom/presto.} pulsar software suite, these
data were partially dedispersed into a reduced number of subbands (generally
16 or 32 subbands) at the average DM of the cluster pulsars before
further timing or search analysis. This process affords an
order-of-magnitude reduction in data size (enabling transfer of these data
over the internet from Puerto Rico to Canada), while still providing the
possibility of creating dedispersed time series at a variety of DMs around
the average DM of the cluster.  These data were also searched for new
pulsars, in the manner described in \S3.

\subsection{Search Sensitivity}

We can estimate the typical minimum flux density to which our searches
were sensitive, as a
function of the radiometer noise and observed pulsar duty cycle, using the
equation 
\begin{equation}
S_{\rm min} = \frac{\sigma \xi T_{\rm sys}}{G\sqrt{n \Delta \nu T_{\rm obs}}}
\left(\frac{w_{\rm obs}}{P_{\rm spin}-w_{\rm obs}}\right)^{1/2}
\label{sens.eqn}
\end{equation}
\citep[following][]{dtws85}.  The minimum signal-to-noise ratio (S/N) of a
search candidate is indicated by $\sigma$, and is taken to be 10 here
(although when candidate lists were short, due to a lack of RFI, we
investigated candidates below this threshold).  $\xi$ is a factor that
incorporates both losses due to the 3-level quantization of the signal and
other systemic effects.  Zero lag and van Vleck corrections \citep[see][and
references therein]{lk04} have been applied to our data.  We use a value
$\xi = 1.2$ to quantify the loss in sensitivity compared with infinite
quantization.  $T_{\rm sys}$ is the equivalent temperature of the observing
system and sky (approximately 40\,K toward our sources, which are
predominantly at high Galactic latitudes). $G$ is the telescope gain, which
is a function of zenith angle, and is taken to have an average value of
10.5\,K/Jy.  For clusters that are only visible at zenith angles $\gtrsim
16^{\circ}$, the gain (and hence the sensitivity) is reduced by 5$-$15\%.
This is the case for the clusters which are closest to the declination limit
of the Arecibo-visible sky: M2, M5, M13, Pal~5, Pal~15, NGC~6535, NGC~6749, and
NGC~6760. $n$ is the number of orthogonal polarizations that have been summed
($n = 2$ here).  $\Delta \nu$ is the bandwidth of the backend, taken to be
100\,MHz here. For clusters where subsequent timing observations were also
performed (M3, M5, M13, M71, and NGC~6749), we were able to search a larger
bandwidth ($\sim 250$\,MHz) by combining multiple WAPPs.  $T_{\rm obs}$ is
the integration time, which varies between clusters from 0.6$-$2.8\,hr
depending on the declination of the source, but is set to 2\,hr for the
purposes of these sensitivity calculations. $w_{\rm obs}$ is the observed
pulse width, which is a function of the intrinsic pulse width $w_{\rm int}$
and other effects that smear the observed pulse profile
(Equation~\ref{obswidth.eqn}). $P_{\rm spin}$ is the pulsar spin period.
 
When the DM of a cluster is not known, one must construct hundreds of trial
time series at a wide range of DMs, each of which must be searched.  Even
for clusters with known DMs, the pulsars have a spread in DM, $\Delta \rm{DM}$,
and several trial DMs must be searched in order to maintain maximum
sensitivity. $\Delta \rm{DM}$ increases roughly linearly with DM, and
typically $\Delta \rm{DM} / \rm{DM}$ is a few percent \citep{fhn+05}.  

The observed pulse width, $w_{\rm obs}$ in Equation~\ref{sens.eqn}, is
always equal to or larger than the pulsar's intrinsic pulse width $w_{\rm
int}$.  Broadening is due to the finite time sampling of the data recorder,
$t_{\rm samp}$, dispersive smearing across individual frequency channels,
$t_{\rm DM}$, smearing due to the deviation of a pulsar's true DM from the
nominal DM of the time series used for searching or folding, $t_{\Delta {\rm
DM}}$, and interstellar scattering, $t_{\rm scatt}$.  One can express the
observed width as the sum in quadrature of these terms:

\begin{equation}
{w_{\rm obs}}^2 = {w_{\rm int}}^2 + {t_{\rm samp}}^2 + {t_{\rm DM}}^2 + {t_{\Delta {\rm
DM}}}^2 + {t_{\rm scatt}}^2,
\label{obswidth.eqn}
\end{equation}

\noindent where the dispersive smearing (assuming the channel bandwidth
$\Delta\nu_{\rm chan} \ll \nu_{\rm center}$) across an individual
channel is given by

\begin{equation}
t_{\rm DM} = 8.3\left(\frac{{\rm DM}}{{\rm
pc~cm^{-3}}}\right)\left(\frac{\Delta\nu_{\rm chan}}{{\rm MHz}}\right)\left(\frac{\nu_{\rm center}}{{\rm GHz}}\right)^{-3}\mu{\rm s},
\end{equation}

\noindent the smearing due to an incorrect DM in the time series is

\begin{equation}
t_{\Delta{\rm DM}} = 4.1\left[\left(\frac{\nu_{\rm low}}{{\rm GHz}}\right)^{-2} -
\left(\frac{\nu_{\rm high}}{{\rm GHz}}\right)^{-2}\right] \left(\frac{\Delta{\rm
DM}}{{\rm pc~cm^{-3}}}\right){\rm ms},
\end{equation}

\noindent (where $\nu_{\rm low}$ and $\nu_{\rm high}$ are the low and high
frequency edges of the bandwidth respectively), and the scattering
($t_{scatt}$ is in ms) can be estimated by the empirical formula
\citep{bcc+04}

\begin{equation}
%Cordes and Lazio 03 version
%{\rm log}_{10}(t_{\rm scatt}) = -3.59 + 0.129~{\rm log}_{10}({\rm DM}) +
%1.02~({\rm log}_{10}({\rm DM}))^{2.0} -
%4.4~{\rm log}_{10}\left(\frac{\nu_{\rm center}}{{\rm GHz}}\right)\,\rm{\mu s}.
{\rm log}_{10}(t_{\rm scatt}) = -6.46 + 0.154~{\rm log}_{10}({\rm DM}) +
1.07~({\rm log}_{10}({\rm DM}))^{2} -
3.86~{\rm log}_{10}\left(\frac{\nu_{\rm center}}{{\rm GHz}}\right).
\label{scatt.eqn}
\end{equation}

For these data at 1.4\,GHz, $t_{\rm DM} = 30-$120\,$\mu$s for DMs
25$-$200\,pc cm$^{-3}$ (using $\Delta\nu_{\rm chan}$ = 100/256\,MHz for DM
$< 100$\,pc cm$^{-3}$ and $\Delta\nu_{\rm chan}$ = 100/512\,MHz for DM $>
100$\,pc cm$^{-3}$). In searches where the cluster DM was not known, we
created dedispersed time series with DMs spaced by 1.0\,pc cm$^{-3}$. For
clusters with known DMs, we typically used a spacing of 0.5\,pc cm$^{-3}$
or finer.  Hence, the maximum DM deviation between a pulsar's true DM and
that assumed in making the trial time series is 0.5\,pc cm$^{-3}$, resulting
in a maximum smearing $t_{\Delta {\rm DM}} = 150$\,$\mu$s.  We have estimated
$t_{\rm scatt}$ using Equation~\ref{scatt.eqn} \citep{bcc+04}, which predicts
the scattering time based on the known (or predicted) DM and the observing
frequency. $t_{\rm scatt}$ varies from source to source, but given the
relatively high observing frequency of these data and the high Galactic
latitudes of most of the sources observed here (most of which have
relatively low known or predicted DMs), it does not significantly increase
the smearing already present in the data.

Figure~\ref{sensitivity.fig} shows the search sensitivity determined from
Equations~1$-$5 as a function of DM and period.\footnote{The degradation in
sensitivity to binary pulsars caused by their orbital motion is {\it not}
included in these estimates.  We discuss our sensitivity to binary pulsars
in \S5.1.} We compare the sensitivity of our survey with that of the only
other major survey of these clusters with Arecibo \citep{and93}.  As the
\citet{and93} survey was conducted at 430\,MHz, we use a typical pulsar
spectral index of $-1.8$ \citep{mkkw00spec} as well as a flatter spectral
index of $-1.3$ to scale that survey to 1400\,MHz, so that the sensitivities
of both surveys can be more directly compared.\footnote{\citet{mkkw00spec}
find a mean value for spectral index, $\alpha$, of $-1.8 \pm 0.2$.  It has
been shown by a number of authors \citep[e.g.][]{kxl+98,tbms98} that the
mean spectral indices of MSPs and un-recycled pulsars are consistent with
each other. Though radio pulsars, both MSPs and un-recycled, generally have
steep, power-law spectra, the observed range of spectral indices in the
pulsar population is large: $0 \gtrsim \alpha \gtrsim -4$.} For low-DM
pulsars with spin periods $\gtrsim 10$\,ms and a standard spectral index
($\alpha = -1.8$), the \citet{and93} survey had a similar sensitivity to
ours. However, for pulsars spinning faster than this, especially those at
high DMs and with relatively flat spectral indices, our survey provides a
significant increase in sensitivity. For example, we were roughly 3$-$6
times more sensitive to a 2-ms pulsar at a DM of 100\,pc cm$^{-3}$, assuming
$-1.8 < \alpha < -1.3$.

Our searches are likely the deepest searches for GC pulsars yet undertaken.
Though recent searches with Parkes \citep[e.g.][]{clf+00} and GBT
\citep[e.g.][]{rhs+05b} benefit from larger recordable bandwidth and longer
possible source tracking times, Arecibo's much larger gain still more than
compensates for these (Table~\ref{surveys_comp.tab}).  Comparing raw gains,
maximum tracking times, and recordable bandwidths, simple scaling using
Equation 1 and a spectral index of $-1.8$ to compare different observing
frequencies, shows that, with the currently available data-recorders,
Arecibo has a raw sensitivity $4 \times$ greater than Parkes and $2 \times$
greater than the GBT for an isolated pulsar.  For binary pulsars, where
blind search sensitivity doesn't improve as $T_{\rm obs}^{1/2}$ (see \S3.2 and
5.1.1.), and longer tracking times provide less benefit, Arecibo is more
sensitive by an additional factor of roughly 2.

\section{Analysis}

\subsection{Radio Frequency Interference Excision}

Periodicities due to terrestrial radio frequency interference (RFI) can
swamp search candidate lists.  Ultimately this reduces a survey's
sensitivity by increasing the number of false positives.  First, strong
bursts of interference, principally from airport and military radars, were
removed in the time domain by clipping samples found to be further than
6\,$\sigma$ from the mean in the ${\rm DM} = 0$\,pc cm$^{-3}$ time series.
Secondly, we created time-frequency masks and applied them to all data
before searching. These masks remove certain channels during specific time
intervals when either the maximum Fourier power, standard deviation, or mean
of the data surpass statistically determined thresholds. This method was
very useful at excising strong narrow-band and transient RFI from the data
before searching and typically only $\lapp 10$\% of the data were masked.
Lastly, we removed known ``birdies'' (weak but highly periodic broadband
interference and bright known pulsars) from the power spectrum by setting
the powers in these narrow frequency intervals to zero. We found that using
these techniques and observing in the relatively RFI-clean frequency ranges
of 1120$-$1220\,MHz and 1370$-$1570\,MHz kept the size of our candidate
lists manageable, while minimizing the risk of rejecting potentially
interesting candidates.

\subsection{Search Techniques}

In a standard Fourier-based search, which identifies pulsar harmonics above
a certain threshold, binary pulsars are strongly selected against because of
orbital modulation, which smears the signal power over many bins in
frequency space.  As the majority of MSPs in GCs are in binary systems, 
sometimes with orbital periods comparable to or shorter than the observation
length, it is crucial to use more sophisticated techniques. 

Our primary search method is a Fourier-based matched filtering technique
\citep{rem02} which assumes that the pulsar's orbit can be described by a
constant acceleration (i.e. constant frequency derivative) over the course
of the observation.  This technique is a frequency-domain version of a
previously used constant acceleration technique \citep{mk84,agk+90,clf+00}
in which the time series is quadratically stretched or compressed before it
is Fourier transformed, in order to simulate different accelerations.  The
advantage of the Fourier-based technique used here is that it is
computationally more efficient than the equivalent time-domain-based method,
and provides even sampling of frequency derivative space. This method is
typically most sensitive to binaries where the orbital period is $\gapp 10\times$
the integration time of the observation, although very bright pulsars with
orbital periods much shorter than this can often be detected \citep{jk91}.
As our Arecibo observations were typically from 1$-$2.5\,hr in length, for
full tracks this search method is most sensitive to orbital periods $\gapp
10-$25\,hr. Using the same technique, we also searched overlapping
subsections of the data sets (which were typically about one third the total
observation length), in order to be sensitive to larger accelerations and
tighter binaries ($P_{\rm orb} \gtrsim 3-$10\,hr), as well as eclipsing
systems.

To search for binaries with orbital periods a factor of $\sim 2$ or more
shorter than the observation time, we used a ``phase modulation'' search
\citep{rce03,jrs+02}. This method makes use of the change in phase of the
pulsations from a binary pulsar and the ``comb'' pattern created by such a
signal in the Fourier power spectrum.  By taking Fast Fourier Transforms
(FFTs) of short subsections of the power spectrum from the full observation,
one can detect the sidebands, which are evenly spaced at the orbital period
and centered on the true frequency of the pulsar.  This method is
appropriate for binary pulsars with orbital periods less than half the
observation length, and increases in sensitivity as the ratio $T_{\rm
obs}/P_{\rm orb}$ increases.  Given the 1$-$2.5\,hr integration times used
in this survey, this technique was sensitive to an area of orbital parameter
space in which no binary radio MSPs have yet been observed.

\subsection{Data Analysis Pipeline}

Here we summarize the data analysis pipeline applied to the observations of
each cluster.  The routines used in the pipeline are part of the PRESTO
pulsar search and analysis package.  First, we generated an RFI mask from
the raw data (as described in \S3.1). Applying this mask, a DM = 0\,pc
cm$^{-3}$, topocentric time series was created and the corresponding power
spectrum was examined by eye for periodic interference.  We added the worst
of the interference to the ``birdie list'' of frequency intervals to be
subsequently removed from the Fourier transform.  Again applying the RFI
mask, we then dedispersed the data at a number of trial DMs using DM steps
$\leq 1$\,pc cm$^{-3}$, clipping samples found to be greater than
6\,$\sigma$ from the mean in the DM = 0\,pc cm$^{-3}$ time series.  The time
series were transformed to the solar-system barycenter during dedispersion,
which allowed us to compare directly candidate periods from separate
observations.  This was very useful for distinguishing likely pulsar
periodicities from the RFI background, which can vary greatly between
observations.  For clusters with known pulsars, we dedispersed at a range of
trial DMs equal to at least 10\% of the cluster's average DM.  For clusters
with no known pulsars, we created time series at a range of DMs centered
around the predicted DM of the cluster \citep{cl02spec}, given its Galactic
coordinates and distance in the \citet{har96} catalog.  We assumed at least
a 100\% error in the predicted DM value when choosing a DM search range.
For example, for a cluster with a predicted DM of 100\,pc cm$^{-3}$, we
searched dedispersed time series with ${\rm DM} = 0-$200\,pc cm$^{-3}$.  The
dedispersion and subsequent analysis of the time series was conducted in
parallel using multiple processors on a 52-node dual-processor Linux cluster
called ``The Borg'', located at McGill University and constructed by our
group specifically for pulsar searches. Once the dedispersion was complete,
the time series were Fourier-transformed. Our analysis did not restrict the
number of samples in the time series to be a power of two.\footnote{Some
minimal padding was added to the data-sets however, so that the number of
samples could be factored into primes where the maximum factor size was
$\leq 13$.} We then set the frequency intervals of the birdie list to zero
in the power spectra before searching the power spectra with both the
phase-modulation and matched-filtering techniques described in \S3.2.  

As most pulsars have relatively short duty cycles, with spectral power
divided between numerous harmonics, we summed harmonics in our matched-filter
search to increase sensitivity to such signals.  For each candidate signal,
sums of 1, 2, 4, and 8 harmonics were tried and the optimum combination was
determined.  In these searches, we looked for signals where the highest of
the harmonics used in summing drifted by up to $z_{\rm max}$ = 170 bins in
the Fourier domain. Higher order harmonics will drift by $N_{\rm harm}$
times more bins than the fundamental, where $N_{\rm harm} = 1$ for the
fundamental, $N_{\rm harm} = 2$ for the second harmonic, etc. If, for
example, 8 harmonics were summed in the identification of a particular
candidate period, then the maximum number of bins the fundamental could
drift by during the observation and still be detectable by our search would
be 170/8. Conversely, for signals where the fundamental drifted by more than
170/2 bins during the observation, we were sensitive to at most one
harmonic. The maximum number of bins a signal is allowed to drift
corresponds to a maximum line-of-sight acceleration of $a_{\rm max} = z_{\rm
max} c P_{\rm spin} T^{-2}_{\rm obs} N^{-1}_{\rm harm} \simeq 4 P_{\rm spin,ms}
 T^{-2}_{\rm obs,h} N^{-1}_{\rm harm}$\,m s$^{-2}$, where $P_{\rm spin,ms}$ and
$T_{\rm obs,h}$ are the spin period in milliseconds and the observation
time in hours and $N_{\rm harm}$ is the highest order harmonic used in
summing. Overlapping subsections of the observation, corresponding to about
a third of the total observation length, were also searched using the
matched-filtering technique and $z_{\rm max}$ = 170 bins, in order to look
for more highly accelerated pulsars. 

\begin{figure}
\centerline{\includegraphics[height=1.0\columnwidth,angle=-90]{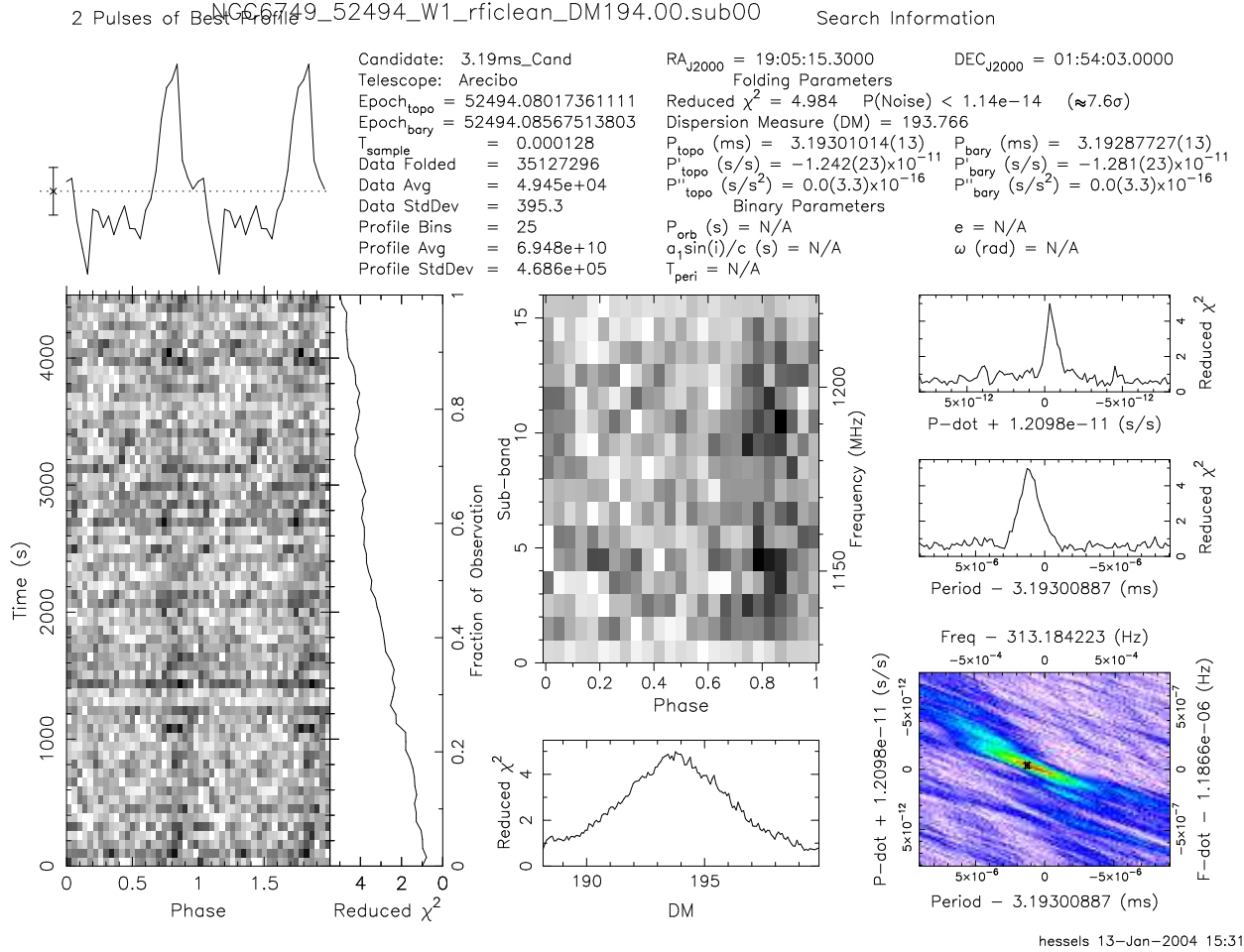}}
%\plotone{f2.eps}
\caption{Sample candidate plot showing the various criteria on which a 
candidate is judged.  The left-most panel shows the pulse intensity as a
function of observing time and pulse phase (two cycles plotted) in
greyscale, with the cumulative profile plotted at top and a side-bar at
right showing the increase in the reduced ${\chi}^2$ (measure of
signal-to-noise ratio) with observing time.  The top-middle greyscale panel
shows the signal strength as a function of pulse phase and observing
frequency (subband).  The bottom-middle plot is the reduced ${\chi}^2$ as a function of DM.
The top-right (top-middle) panel shows the reduced ${\chi}^2$ as a function
of trial period derivative (period) at the best period (period derivative).  Finally, the
bottom-right panel shows the reduced ${\chi}^2$ over a range of trial periods
 and period derivatives.  During the folding, a range of DM, period,
and period derivative are searched to produce the highest reduced ${\chi}^2$.
This optimization improves on candidate parameters determined during the initial search.
This is the discovery  observation of NGC~6749A.  
The subbands show effects due to the masking of some 
of their constituent channels to remove RFI.
\label{discplot.fig}}
\end{figure}

The resulting candidate lists from the matched-filtering search were
generally short enough that they could easily be examined by eye, although
we also parsed candidate lists with a script that automatically folded
candidates above an equivalent gaussian significance threshold of 7.
Interesting candidates were folded using the estimated DM, period, and
period derivative from the search and these parameters were optimized by the
folding software to maximize the S/N ratio of the folded
profile.  The output plot from such a fold was used to determine whether a
given candidate warranted further attention (see sample discovery plot and
description in Figure~\ref{discplot.fig}).  To identify potentially
interesting candidates from the phase-modulation searches, we compared the
search outputs from the different observing epochs of each cluster. The
criteria for a promising phase-modulation candidate was a signal with a
significance $> 10\sigma$ that did not peak in significance at DM $= 0$\,pc
cm$^{-3}$, had an orbital period $> 600$\,s, and appeared, by virtue of a
similar orbital period, in the candidate lists of at least two observing
epochs. 

\section{Results}

\subsection{Redetections}

Table~\ref{prev_pulsars.tab} lists the 15 previously known pulsars in our
survey clusters.  The majority of these pulsars were easily detected by our
search pipeline (Table~\ref{prev_pulsars.tab}).  In fact, because many of
these sources are relatively bright, masking these periodicities and their
many significant harmonics was an important factor in reducing the length of
candidate lists.  The only previously known pulsars not detected in our
searches are M15F, G, and H.\footnote{Note however that M15F was easily
detected in complementary searches we made using the WAPP at 327\,MHz (see
\S4.2.2).} Although these are all isolated pulsars, and are well within
the $\sim 1.5^{\prime}$ half-power radius of the 1.4-GHz Arecibo main beam,
our non-detections are not very surprising: these are the dimmest pulsars in
M15, and were all found in searches of multiple, combined observing epochs
\citep{and93}. Assuming standard spectral indices ($\alpha = -1.8$), they
have flux densities right at the limit of our search sensitivity.

\subsection{New MSPs}

\begin{figure}
\centerline{\includegraphics[height=1.0\columnwidth,angle=-90]{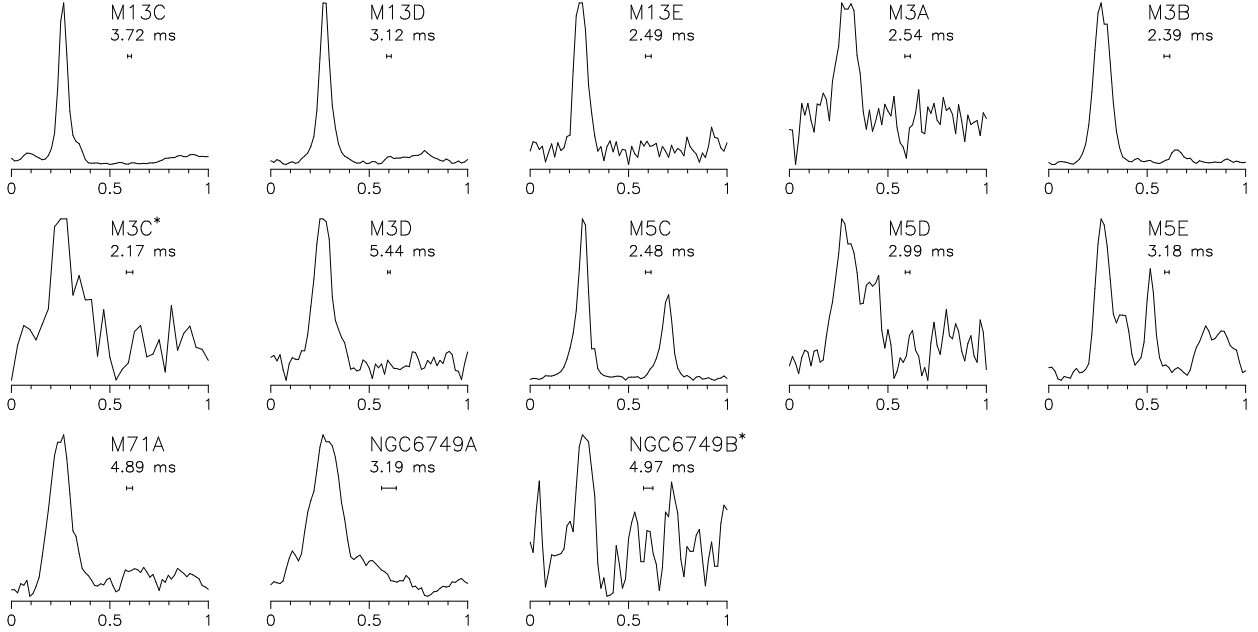}}
%\plotone{f3.eps}
\caption{$1.4$-GHz pulse profiles for the 11 millisecond pulsars and two
promising candidates (marked with an asterisk) discovered in this survey.
The profiles are often the sum of numerous observations and in the case of
clusters where there is significant scintillation (i.e. M3, M5, and M13) the
profiles are the sum of a few (or only one) observations where the pulsar
appears much brighter than on average.  The profiles have been rotated in
phase so that pulse maximum occurs at 0.25.  In each case, there are 64 bins
accross the profile, with the exception of M3C where 32 bins are used.  The
horizontal bar indicates the effective time resolution of the data, taking
into consideration dispersive smearing.  To aid the reader in interpreting
what features in these profiles correspond to real emission, rather than
spurious baseline fluctuations, we identify the following.  M13C has low
level emission preceding the main pulse and starting at phase 0.7.  M13D may
have a very weak interpulse around phase 0.75.  M3B has a weak interpulse
around phase 0.65.  M5C has an interpulse around phase 0.7, which is of
comparable strength to the main pulse.  M5E has a duty cycle close to 100\%,
showing at least 3 significant peaks.  \label{profiles.fig}}
\end{figure}

We have discovered 11 MSPs and 2 promising candidates in 5 clusters. Three
of the clusters with new pulsars (M3, M71, and NGC~6749) contained no known
pulsars prior to our survey.  All of the pulsars discovered in this survey
were found using the matched-filtering acceleration search technique, and
all but the newly-found isolated pulsar M13C (\mxiiic) and the
long-orbital-period binary M3D (\miiid) required a non-zero trial frequency
derivative (acceleration) in order to be detectable.  Given the criteria
outlined in \S3.3, no interesting candidates were identified by the
phase-modulation search. The spin periods of the new pulsars have a narrow
range 2.4$-$5.4\,ms. For comparison, the previously known pulsars in these
clusters have spin periods in the range 3.5$-$111\,ms, with only 2 pulsars
having $P_{\rm spin} < 4$\,ms. All but one of the new pulsars is in a binary
system, with orbital periods ranging from $2.1$\,hr up to $129$\,d.
Although none of the previously known binaries in these GCs show eclipses, 3
of the new pulsars found here eclipse. The basic characteristics of these
pulsars are summarized in Table~\ref{pulsars.tab}. Integrated pulse
profiles, which are often the sum of numerous observations, are shown in
Figure~\ref{profiles.fig}.

We have conducted monthly timing observations of these discoveries over the
course of approximately two years using Arecibo and multiple WAPP backends.
In fact, M3D (\miiid), M5E (\mve), and M13E (\mxiiie) were all discovered in
searches of timing data because of fortuitous scintillation.  The timing
results for the M5 and M71 pulsars will be presented by Stairs et al. (2008,
in preparation) and the timing of the M13 pulsars will be presented by
Ransom et al. (2008, in preparation).

\subsubsection{M3 (NGC~5272)}

We have found the first three, and likely four, pulsars known in M3. All of
these pulsars were detected in observations in which interstellar
scintillation increased their flux to a detectable level and they are not
consistently detectable with Arecibo.  They have DMs within 0.3\,pc
cm$^{-3}$ of each other, with an average DM of 26.4\,pc cm$^{-3}$.  This
compares very well with the 23\,pc cm$^{-3}$ DM predicted by the NE2001
model \citep{cl02spec}.  On these grounds alone, there is little doubt that
these pulsars are members of M3. M3A (\miiia) is a 2.54-ms binary which has
only been detected three times, on MJDs 52491, 52492 and 52770 (we list the
number of detections of each new pulsar in Table~\ref{pulsars.tab}).  Due to
this paucity of detections, we do not currently know the orbital parameters
for this pulsar, although its orbital period is likely on the order of a
day. We also cannot derive a precise position for this pulsar, though we
note that, since the half-power radius of the Arecibo 1.4-GHz main beam is
$\sim 1.5^{\prime}$, it is likely no further than this from the cluster
center. M3B (\miiib) is a 2.39-ms binary in a 34.0-hr orbit with a
0.2\,\msun\ (minimum mass) companion, which, in analogy to other systems
with similar orbital parameters, may be a low-mass helium white dwarf. It is
the most consistently detectable pulsar in M3, and is visible at least
faintly in roughly half of our observations.  On some occasions M3B has
shown very large increases in flux (up to a flux density at 1400\,MHz
$S_{1400} \sim 0.1$\,mJy), presumably due to diffractive scintillation.  We
have derived a phase-connected timing solution for M3B
(Table~\ref{M3pulsars.tab}) using the {\tt TEMPO} pulsar timing
package\footnote{http://www.atnf.csiro.au/research/pulsar/tempo} and
standard pulsar timing techniques.  This solution places it $8.3^{\prime
\prime}$ (0.25 core radii) from the center of the cluster.  Using this
projected position and the simple cluster model outlined in \citet{fhn+05},
we can derive the maximum contribution to the observed period derivative
from acceleration in the cluster's gravitational potential.  In this model,
we find $|a_{\rm max}/c|P_{\rm spin} = 1.6 \times 10^{-20}$\,s/s, which is
comparable to M3B's observed period derivative. From this we place an upper
limit on M3B's {\it intrinsic} period derivative and corresponding lower
limits on its characteristic age and dipole magnetic field
(Table~\ref{M3pulsars.tab}).
%The NE2001 model predicts a
%scintillation time scale of 1300\,s (at 1-GHz observing frequency and
%assuming a 100\,km s$^{-1}$ transverse velocity) and a scintillation
%bandwidth of 3\,MHz (also at 1-GHz observing frequency), given the 10.4\,kpc
%distance to M3.  These are significantly smaller than the observation time
%and bandwidth we used, confirming that scintillation is important for this
%cluster. 

M3D has a spin period of 5.44\,ms and an orbital period $\sim 129$\,d.
Though the detections of M3D are too sparse to derive a phase-connected
timing solution for this pulsar, by inserting arbitrary phase jumps between
observing epochs we have been able to derive accurate orbital parameters,
and a reasonably precise position (Table~\ref{M3pulsars.tab}). M3D's orbital
period is much longer than the typical orbital period of GC MSPs, most of
which have $P_{\rm orb} < 3$\,d, and may suggest a non-standard
evolutionary history for the system. PSR~B1310+18 in M53 \citep[$P_{\rm orb}
\sim 256$\,d,][]{kapw91}, PSR~B1620$-$26 in M4 \citep[$P_{\rm orb} \sim
191$\,d,][]{lbb+88}, and PSR~J1748$-$2446E in Terzan~5 \citep[$P_{\rm orb}
\sim 60$\,d,][]{rhs+05b} are the only other GC MSPs known to have orbital
periods longer than 50\,d. Binaries with orbital periods $10 \lapp P_{\rm
orb} \lapp 1000$\,d may be efficiently formed by exchange interactions
involving an isolated neutron star and ``hard'' primordial binaries
\citep{hmg+92,sp93}.  \citet{cr05} point out that pulsars with long orbital
periods ($> 100$\,d) tend to reside in low-density clusters, i.e.
$\rho_\circ < 4$\,log$_{10}$~\lsun/pc$^{3}$.  As M3 has a central density
$\rho_\circ = 3.51$\,log$_{10}$~\lsun/pc$^{3}$, M3D follows this trend.
M3D also has a significant eccentricity, $e = 0.075$.  This is significantly
larger than expected from a single-stage stable-mass transfer episode from a
red-giant \citep{phi92b}, further pointing to an unusual evolution for this
system.
%Cut?
%However, such wide binaries also have large
%interaction cross-sections and may easily be disrupted in the high-density
%cores of GCs.

We also have one very good candidate (denoted M3C in Table~\ref{pulsars.tab}
because it was found before M3D and has been presented as such in previous
references, e.g. \citet{rhs+05a}) with a spin period of 2.17\,ms that has
been seen only once (presumably due to scintillation) with a S/N of $\sim 6$
during a $\sim 3000$-s portion of an observation with the Green Bank
Telescope (GBT), but never in any of our Arecibo data.  During this single
detection, the candidate showed a period drift of $1.82(3) \times
10^{-11}$\,s/s, corresponding to a line-of-sight acceleration $a_l =
2.5$\,m/s$^{-2}$, indicating that, if it is real, this pulsar is in a binary
system.
%Note: Pdot = a_l P_o / c
These GBT data were
taken with the Berkeley-Caltech Pulsar Machine (BCPM) as part of a parallel
survey by our group for pulsars in GCs visible with the GBT \citep[see][for
more details on these survey observations]{rsb+04,rhs+05a}.

A deep radio synthesis image of M3 made at 1.4\,GHz with the VLA
\citep{kgwm90} revealed a $S_{1400} \sim 180$\,$\mu$Jy source
$7.4^{\prime \prime}$ from the optical center of the cluster.  This source
does not coincide with the positions of M3B or M3D. In principle, the
\citet{kgwm90} radio source could be associated with M3A, M3C, or another
unknown pulsar in the cluster.  However, it seems too bright to be M3A or
M3C, and unless it is particularly fast-spinning ($P_{\rm spin} \lesssim
1$\,ms) or highly accelerated by a companion star, its flux density should
have made it easily detectable in our searches.  It is of course also
possible that the source is not associated with M3.
%A timing position is needed in
%order to test the association. We have tried setting the position of M3B
%in our timing model to that of the \citet{kgwm90} source, but still find
%that the detections are too infrequent to allow for phase connection.
Three 10-ks observations of the cluster with {\it Chandra} ACIS-S taken by
Grindlay et al. (ObsIDs 4542, 4543, and 4544) reveal no obvious X-ray
counterpart to the \citet{kgwm90} radio source.  There are however two
obvious point sources within the half-mass radius of the cluster.  One of
these sources is coincident with the supersoft X-ray source 1E 1339.8+2837
\citep*{dag99}.  The other is not coincident with the positions of either
M3B or M3D.
%, which is likely a
%XXX.  The spectrum of the other point source suggest it is a XXX.

\subsubsection{M5 (NGC~5904)}

In M5, we have found three new pulsars in addition to the isolated 5.55-ms
pulsar M5A (\mva) and the binary 7.95-ms pulsar M5B (\mvb) found by
\citet{wak+89a}, bringing the total population of this cluster to five. M5C
(\mvc), with a spin period of 2.48\,ms, is in a 2.1-hr orbit with a
0.04\,\msun\ (minimum mass) companion.  It shows regular eclipses for $\sim
15$\% of its orbit as well as eclipse delays at eclipse ingress and egress,
which can be up to $\sim 0.2$\,ms and are presumably due to dispersive
delays as the pulsar passes through the ionized wind of its companion. M5C
is part of the growing class \citep{fre05} of $\sim 10$ eclipsing GC
binaries with orbital periods of only a few hours and very low mass
companions ($M_c \lapp 0.1$\,\msun ).  It is positionally coincident with a
soft X-ray counterpart seen in a 45-ks {\it Chandra} ACIS-S observation of
the cluster (Stairs et al. 2008, in preparation).  M5D (\mvd), a binary
2.99-ms pulsar, was originally discovered in Arecibo data taken by our group
at 327\,MHz \citep[see also][]{mf03}, but it has also been seen at 1.4\,GHz
on numerous occasions because of scintillation.  These 327-MHz data were
obtained using the Gregorian 327-MHz receiver and the WAPP as part of a
smaller set of search observations conducted at lower frequency on the
clusters M3, M5, M13, and M15.  M5D is in a 29.3\,hr orbit with a
0.20\,\msun\ (minimum mass) companion.  M5E, a binary 3.18-ms pulsar, was
discovered in a search of our regular timing observations of M5 and was
visible because of scintillation.  M5E has an orbital period of 26.3\,hr and
a 0.15\,\msun\ (minimum mass) companion. It has a complex pulse profile and
close to a 100\% duty cycle (Figure~\ref{profiles.fig}).

\subsubsection{M13 (NGC~6205)}

In M13, we have found three new pulsars in addition to the isolated 10.4-ms
pulsar M13A (\mxiiia) and the binary 3.53-ms pulsar M13B (\mxiiib) found by
\citet{kapw91} and \citet{and93}, bringing the total population of this
cluster to five. The pulsars in this cluster show 2$-10\times$ changes in
flux density because of scintillation on time scales shorter than an hour.
M13C has a spin period of 3.72\,ms and is the only isolated pulsar
discovered in this survey.  M13D (\mxiiid) is a 3.12-ms binary with a
14.2-hr orbital period and a 0.18\,\msun\ (minimum mass) companion. M13E, a
binary 2.49-ms pulsar, has been detected in only two observations, likely
because of favorable scintillation.  It is highly accelerated and appears to
be eclipsed for part of each of these two observations.  We estimate that
the orbital period is approximately $2.8 \pm 0.2$\,hr, which is consistent
with the interpretation that it is similar to M5C.  The short orbital period
and likely eclipses of this system make it difficult to blindly detect in a
search.  In analogy to other eclipsing MSPs, it is also likely that M13E
will be visible as an X-ray source.  Two roughly 30-ks {\it Chandra}
observations of M13 were taken in March 2006, and may reveal an X-ray source
coincident with this pulsar, or one of the others known in M13.

\subsubsection{M71 (NGC~6838)}

In M71, we have found M71A (\mlxxia), the first and only pulsar known in
this cluster.  M71A's DM of 117 pc cm$^{-3}$ is reasonably close to the
$86$\,pc cm$^{-3}$ predicted by the NE2001 Galactic electron model
\citep{cl02spec}.  It is also located $\sim 0.6$ core radii from the optical
center of the cluster (see Stairs et al. 2008, in preparation), leaving
little doubt it is associated with M71.  M71A is a 4.89-ms pulsar in a
4.2-hr orbit with a 0.03\,\msun\ (minimum mass) companion.  It shows regular
eclipses for $\sim 20$\% of its orbit, though no eclipse delays are seen at
ingress or egress.  A discussion of its identification with an X-ray
counterpart will be published elsewhere. The low density and relative
proximity of M71 (d = 4.0\,kpc) make optical follow-up observations of M71A
viable. The relatively high DM towards this cluster means diffractive
scintillation has a small effect on the flux density of M71A. A stack search
combining four contiguous days of data on this cluster revealed no new
pulsars; searches combining more data sets will be undertaken.
%The lack of visible eclipse delays is likely due to *** XXX ***.

\subsubsection{NGC~6749 (Berkeley~42)}

Lastly, we have found the first known pulsar in NGC~6749, as well as another
promising pulsar candidate in this cluster.  NGC~6749 has the lowest
concentration ($c = {\rm log}_{10}(r_t/r_c)$) and the third lowest central
luminosity density (tied with M13) of any GC with a known pulsar. NGC~6749A
is a 3.19-ms binary pulsar with an orbital period of 19.5\,hr and a
0.090\,\msun\ (minimum mass) companion.  Though there is no evidence for
eclipses in the system, we cannot rule them out, as orbital coverage around
superior conjunction is poor. The DM of NGC~6749A (\nvia) is 194 pc
cm$^{-3}$, which is significantly lower than the 438\,pc cm$^{-3}$ DM
predicted by the NE2001 model \citep{cl02spec}.  Due to the sparseness of
measured arrival times at some epochs, we have only been able to derive a
{\it partially} phase-connected timing solution for this pulsar
(Table~\ref{M3pulsars.tab}), which places the pulsar $0.51 \pm 0.38
^{\prime}$ ($\sim 0.7$ core radii) from the center of the cluster. Thus, an
association between this pulsar and NGC~6749 is quite
secure\footnote{NGC~6749 sits in the Galactic plane at a latitude of
$-2.2^{\circ}$.  Using the number of observed field MSPs in the Galactic
plane to roughly estimate the angular density of observable MSPs, we use
this number and the angular distance of NGC~6749A from the cluster center to
estimate a chance association probability of $\sim 10^{-8}$.}, and the DM
discrepancy could easily be due to uncertainties in the NE2001 model, or
perhaps an overestimation of the cluster's distance.  

NGC~6749B (\nvib) is a candidate 4.97-ms binary pulsar that has been seen
only once, in data from MJD~52921. Although it is quite faint -- the
detection has a S/N of $\sim 5$ -- the signal shows a clear peak in DM at
roughly 192\,pc cm$^{-3}$.  The similarity in DM with NGC~6749A bolsters
this candidate's identification as a real pulsar and member of NGC~6749.  
In the one 5000-s observation where this candidate was seen, it showed a
period drift of $-1.53(2) \times 10^{-11}$\,s/s, corresponding to a
line-of-sight acceleration $a_l = -0.9$\,m s$^{-2}$, and indicating that, if
this is a real pulsar, it is in a binary system.

\section{Discussion}

%MAIN TOPICS:
%1) What is the intrinsic orbital period dist?
%2) What is the intrinsic spin period dist?
%3) What GC have MSPs?
%4) How do the populations of MSPs differ between
%   clusters and between GCs and the Plane?

\subsection{Survey Limitations}

In this section, we discuss the limitations of this survey due to a number
of observational and analytical biases. Before we do, it also bears
reminding the reader that the number of pulsars we can find is also limited
in a more fundamental way by the clusters' efficiency at creating them. From
theoretical expectation and mounting observational evidence, it is becoming
clear that cluster density has an important role to play in creating MSPs in
a GC.  Currently, no cluster with a central luminosity density of
$\rho_{\circ} < 3$\,log$_{10}$~\lsun/pc$^{3}$ contains a known pulsar.
Conversely, Terzan~5 and 47~Tuc have central densities of $\rho_{\circ} =
5.06$ and $4.81$\,log$_{10}$~\lsun/pc$^{3}$ respectively and have the
largest known populations. We have surveyed all 22 GCs visible from Arecibo
and within 70\,kpc without any bias against observing low-density clusters.
Unfortunately, from the point of view of wanting to find as many new pulsars
as possible, these clusters are on average not very dense compared with
clusters in the Galactic bulge, outside of Arecibo's field of view.  Of the
22 clusters in our survey, 8 have $\rho_{\circ} <
3$\,log$_{10}$~\lsun/pc$^{3}$ and still contain no known pulsar. This is not
a great surprise.  Of the remaining 14 clusters, which have densities in a
range where one might expect to find pulsars, 8 have known pulsars.  The
absence of any known pulsar in the remaining 6 survey clusters with
$\rho_{\circ} > 3$\,log$_{10}$~\lsun/pc$^{3}$ can be mostly explained by
their large distances (see also \S5.1.2).

\subsubsection{Sensitivity to Fast and Binary Pulsars}

The fastest-spinning pulsar known is \terad\ in Terzan~5, with a spin period
of 1.396\,ms \citep{hrs+06} and the shortest known orbital period of any
binary MSP is 1.6\,hr \citep[PSR~J0024$-$7204R in 47~Tucanae,][]{clf+00}.
Here we discuss the sensitivity of this survey to very-fast-spinning pulsars
($P_{\rm spin} \sim 0.5-$3.0\,ms) and/or pulsars in very tight orbits
($P_{\rm orb} \sim 0.5-$6.0\,hr).  The discovery of such systems is hampered
by a number of selection effects \citep{jk91,rce03}, which bias the
observed spin and orbital period distributions to longer periods. We discuss
the extent to which the observed spin period distribution of the total
population of MSPs has been affected by selection effects in \S5.2.  

\begin{figure}
\centerline{\includegraphics[height=0.9\columnwidth]{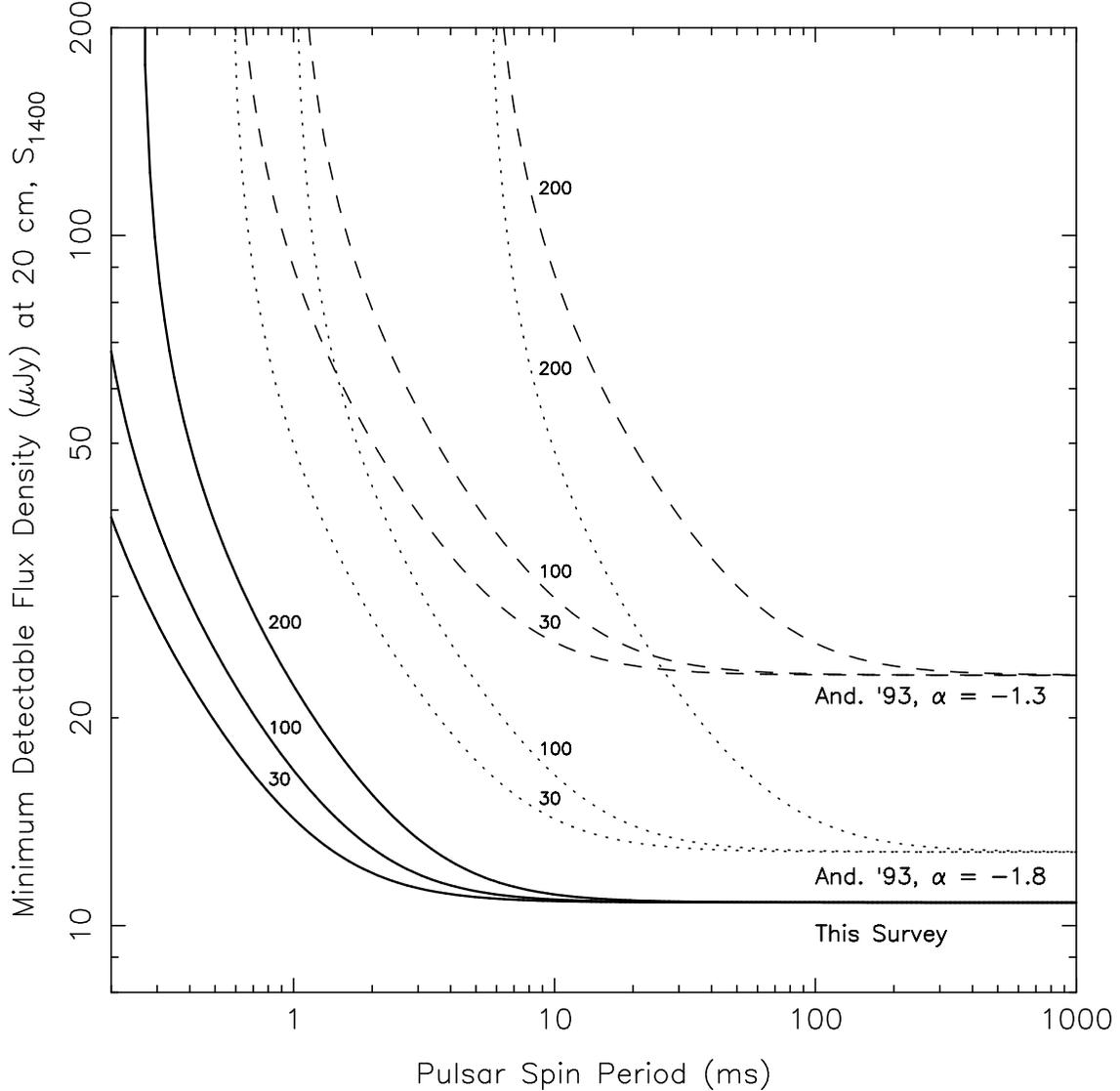}}
%\plotone{f1.eps}
\caption{Survey sensitivity as a function of period and DM,
assuming an intrinsic pulse width of 8\% and an integration time of 2\,hr.
Each set of curves shows (from left to right) the sensitivity for DMs of 30,
100, and 200\,pc cm$^{-3}$.  
%The discontinuities in these curves indicate a
%loss of power in the Fourier domain as higher order harmonics surpass
%half the Nyquist frequency $\nu_{\rm Nyq} = 1/2t_{\rm samp}$.  
The solid curves are the
sensitivity of the survey described in this paper.  The dotted (dashed)
curves are the sensitivity of the 430\,MHz survey of \citet{and93} scaled to
1400\,MHz assuming a spectral index of $-1.8$ ($-1.3$).  For low-DM pulsars with
periods $\gtrsim 10$\,ms and steep spectral indices, the \citet{and93}
survey has comparable sensitivity.  However, the survey presented here is
significantly more sensitive to pulsars with periods $\lesssim 4$\,ms, high
DMs, and/or relatively flat spectral indices.  The plotted survey
sensitivity to slower pulsars ($P_{\rm spin} \gtrsim 100$\,ms) is likely
significantly over-estimated by an unknown factor between $2-10$ due to RFI
and red noise (see \S5.1.1).
\label{sensitivity.fig}}
\end{figure}

We have characterized the sensitivity of this survey, as a function of
period and DM, in \S2.3 and compare it to the sensitivity of the
\citet{and93} survey (Figure~\ref{sensitivity.fig}).  Though the survey's
sensitivity to slow pulsars ($P_{\rm spin} \gtrsim 10$\,ms in this case) is
shown to be flat, there is undoubtedly some extra, unmodeled reduction in
sensitivity to very slow ($P_{\rm spin} \gtrsim 0.5$\,s) pulsars because of
RFI and red-noise. We note however that very few slow pulsars are known in
the GC system and we do not a priori expect such systems in the clusters we
have surveyed because they are generally found in higher-density clusters
\citep{cr05}.

For DM = 30\,pc cm$^{-3}$, our sensitivity to a 1-ms pulsar is degraded by a
factor of roughly $1.5$ compared with a 4-ms pulsar.  This degradation in
sensitivity is a strong function of DM and increases to a factor of $\sim
2.5$ at a DM of 200\,pc cm$^{-3}$. Some of the MSPs we have discovered in
this survey were just barely detectable by our processing (e.g. NGC~6749A
and M3D).  We can thus not rule out the existence of pulsars with $P_{\rm
spin} \lesssim 1.5$\,ms in our survey data, if they have fluxes comparable
to the dimmest sources we have discovered.  However, a ``reasonably bright''
($S_{1400} > 50$\,$\mu$Jy), {\it isolated} 1-ms pulsar would very likely
have been detected.  We discuss the {\it luminosity} limits achieved for
individual clusters, which depend largely on the cluster distance, in the
next section.  Even higher time and frequency resolution data are required
to maintain as flat a sensitivity response as possible out to spin periods
below $\sim 1$\,ms.  This should be a goal of future surveys and is feasible
given the current state of computer technology.

\begin{figure}
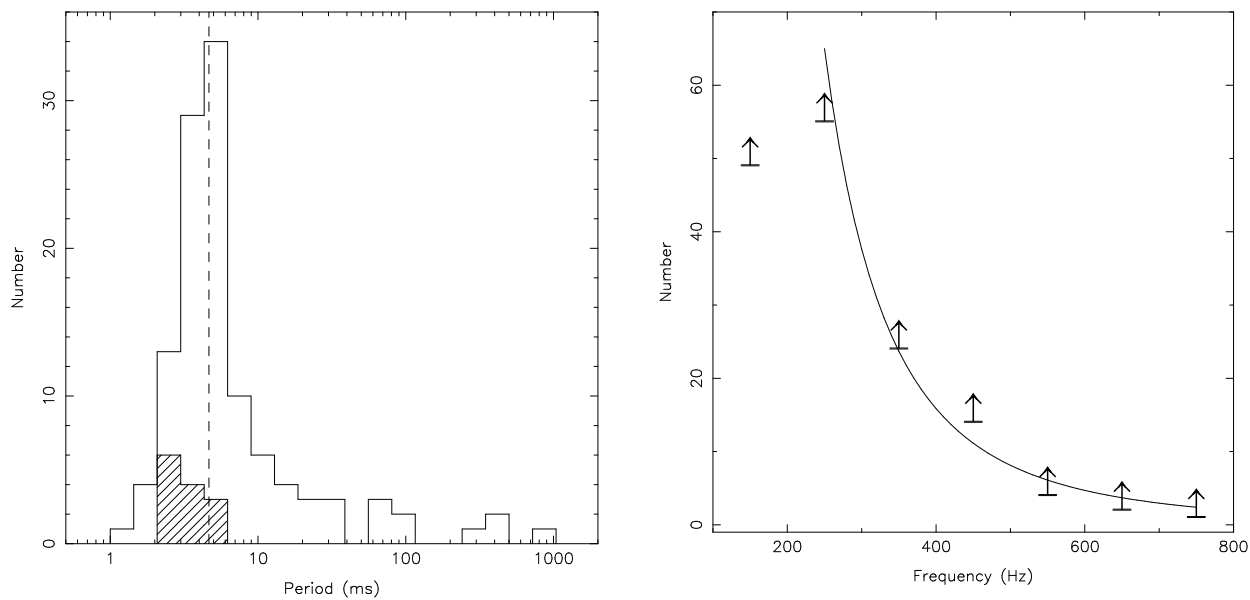

\centerline{\includegraphics[height=0.475\columnwidth,angle=-90]{f4a.eps}
\hspace{0.5cm}
\includegraphics[height=0.475\columnwidth,angle=-90]{f4b.eps}}
%\plotone{f4.eps}
\caption{{\it Left:} Period histogram of 129 known GC pulsars (listed in
http://www.naic.edu/$\sim$pfreire/GCpsr.html).  The shaded area reflects
pulsars found in the survey presented here.  The vertical dashed line marks
the median spin period of the observed population, 4.7\,ms.  {\it Right:} The 
combined population of MSPs in the field and GCs, plotted as a
function of spin frequency.  The points are binned in intervals of 100\,Hz,
and are shown as lower limits to reflect the various observational biases
against detecting millisecond pulsars.
The number of observed pulsars drops rapidly with spin frequency above 200\,Hz,
roughly as $N_{\rm psr} \propto {\nu_{\rm spin}}^{-3}$ (solid line).
\label{spinfdist.fig}}
\end{figure}

The shortest spin period found in this survey was 2.4\,ms (M3B).
Furthermore, 5 of the 11 pulsars found here have spin periods between
2$-$3\,ms, significantly lower than the median period of the observed
population of GC MSPs, which is $4.7$\,ms (see Figure~\ref{spinfdist.fig},
left, for a spin period histogram of all known GC pulsars). For comparison,
the fastest-spinning pulsar known in these clusters prior to this survey is
M13B, with a spin period of 3.5\,ms.  This demonstrates the increased
sensitivity of this survey to fast-spinning pulsars, compared with previous
surveys, and suggests that the observed spin-period distribution of GC
pulsars is still artificially biased towards longer spin periods.  This is
unsurprising if one compares the sensitivity curves of this survey with
those of \citet{and93}, as in Figure~\ref{sensitivity.fig}.

Our sensitivity to short orbital periods is difficult to quantify precisely.
We compare the sensitivity of a coherent search of the data (i.e. one in
which the orbital modulation of the pulsar signal can be completely
corrected, and in which the sensitivity is proportional to
$T^{-1/2}_{\rm obs}$) to that of our acceleration search technique, using
the simulations of \citet{rce03}, which were made using the same search
technique and software. These simulations assume a pulsar spin period of
2\,ms and a companion mass of 0.1\,\msun.  We see that for a 2-hr binary
period and a 0.5-hr observation duration that the sensitivity afforded by an
acceleration search is roughly half that of the coherent sensitivity.  A
0.5-hr observation duration is typical of the length used in our short
subsection searches, and thus for all clusters we had good sensitivity to orbital
periods down to a few hours.  Specifically for the survey clusters with
known pulsars (where the DM is thus also known), we searched a larger variety
of short subsections of the total observation length, down to integration times
as short as roughly 10 minutes.  For these clusters, we were sensitive to
even more compact systems and higher accelerations (assuming the pulsar's
flux is high enough to show up in such short integrations).

As explained in \S3.3, in our matched-filter acceleration technique we
looked for signals where the highest order harmonic used in summing drifted
by up to $z_{\rm max} = 170$ bins, corresponding to a maximum line-of-sight
acceleration of $a_{\rm max} \simeq 4 P_{\rm spin,ms} T^{-2}_{\rm obs,h}
N^{-1}_{\rm harm}$\,m s$^{-2}$.  For comparison, we note that \citet{clf+00} were
sensitive to a maximum acceleration of $< |30|$\,m s$^{-2}$ in their
searches of 47~Tucanae (where 0.3-h integration subsections were used) and
that PSR~B1744$-$24A in Terzan~5 has a maximum line-of-sight acceleration of
33\,m s$^{-2}$ \citep{ljm+90spec}.  For $P_{\rm spin} = 2$\,ms and $T_{\rm obs}
= 0.5$\,h, which is typical of the integration times used in our
short-subsection searches, we reach a limiting acceleration of 32\,m
s$^{-2}$ (note however that this limit applies only to the fundamental and
not to sums including higher-order harmonics).  For clusters where we
searched numerous timing observations, $z_{\rm max} = 500$ bins was
sometimes used, providing an additional factor of three range in
acceleration space.  Such high accelerations are worth exploring in order to
maintain sensitivity to not only the accelerated fundamental of a pulsar
signal but also its harmonics, which are needed to detect faint pulsars.
Though no new pulsars were found in these searches, we note that the pulsars
M13E and M71A were only detected with $1-2$ harmonics in our initial $z_{\rm
max} = 170$ searches and would have been more easily identified in a $z_{\rm
max} = 500$ search.  Re-searching the data presented here with $z_{\rm max} =
1000$ has the potential to discover compact binaries that were previously missed.

The most extreme orbital systems we found in this survey were M5C, M13E, and
M71A, with orbital periods between $\sim 2-$4\,hr and minimum companion
masses between $\sim 0.02-$0.1\,\msun. As 10 of 11 pulsars found here are in
binary systems, this survey did a good job of detecting the binaries that
were missed by \citet{and93} and other previous searches of these
% From Anderson Thesis:
% M15A,B readily detected in normal Fourier techniques
% M15C found in acceleration search
% M15D,E incoherent stack search
% M15F,G,H coherent multi-day search
clusters.\footnote{However, we note that five of the eight pulsars found in
M15 by \citet{and93} are especially faint, isolated pulsars discovered in 
searches of multiple M15 data sets.  M15D and E were found in incoherent
stack searches of multiple observations, while M15F, G, and H were found in
coherent multi-day transforms. These searches are {\it much} less sensitive
to binary systems.}  With the exception of the long orbital period binary
M3D, all the binaries presented here required the acceleration search
technique in order to be detected. The fact that we found only 1 isolated
pulsar suggests that previous surveys already found the vast majority of the
reasonably bright isolated pulsars in these clusters.

%From Camilo \& Rasio Aspen Proc:
%``It is clear by comparison with otherwise equivalent pulsars in the
%Galactic disk that even the ``small'' eccentricities of most GC binaries
%are unusually large - a clear sign of stellar interactions either during
%or post formation''

%Cluster binaries have shorter orbital periods than those in the field.

%Johnston \& Kulkarni 1991 \citep{jk91}

%Camilo et al. 2000 run simulations for 47Tuc

\subsubsection{Sensitivity to the Weakest Pulsars}

%Important to remember that GCs are relatively far away, and that we
%see only the brightest portion of the luminosity distribution.
%Is our Smin low enough to detect the dimnest pulsars seen in Ter5?
%Are we starting to see a cutoff in the luminosity distribution?

Of the 22 clusters we have searched for pulsations, 14 still contain no
known pulsar, although all of these (with the exception of M2) are $>
15$\,kpc from the Sun and/or have very high predicted DMs ($> 150$\,pc
cm$^{-3}$).  We have estimated the maximum luminosity of any undiscovered
pulsars in the clusters we have searched, using the distance to the cluster,
its DM (or predicted DM), and the sensitivity calculations of \S2.3.  These
upper limits are given in Table~\ref{gcs.tab}, and apply most directly to
isolated pulsars.  For comparison, the weakest-known pulsars in Terzan~5
have 1400-MHz luminosities\footnote{This is a ``pseudo'' luminosity, defined
as $L_{1400} \equiv S_{1400} d^{2}$.} $L_{1400} \sim 2$\,mJy
kpc$^2$, and the weakest in 47~Tucanae are $L_{1400} \sim 1$\,mJy
kpc$^2$. Furthermore, there are indications that the {\it intrinsic} lower
limit for the MSPs in 47~Tucanae is $L^{min}_{1400} \approx 0.4$\,mJy kpc$^2$
\citep{mdca04}. Because of their large distances, such weak pulsars are not
excluded in any of the clusters we have searched here. For most of our
clusters, the luminosity limits only exclude pulsars whose luminosity is
comparable to the brightest MSPs known in the GC system ($L_{1400} \gtrsim
5$\,mJy kpc$^2$). Significantly more sensitive observing systems (using,
for example, the Square Kilometer Array) will be required to fully probe the
pulsar populations of these clusters down to the proposed low-luminosity
cutoff.

\subsubsection{Spatial Coverage}

Here we investigate whether our single-pointing observations provided
adequate spatial coverage to discover the bulk of the visible pulsars in our
survey clusters.  The vast majority of GC pulsars are found close to the
centers of their clusters due to mass-segregation of the massive neutron
stars.  Of the GC pulsars with measured angular distance from their host
cluster's center, $r$, roughly 90\% are at $r/r_c < 3$ \citep[see
e.g.][]{cr05}, where $r_c$ is the cluster's core radius (these are listed
for each of our survey clusters in Table~\ref{gcs.tab}).  Pulsars further
from their cluster center ($r/r_c \gtrsim 5$) are found predominantly in
high-density ($\rho_{\circ} \gtrsim 4.5$\,log$_{10}$~\lsun/pc$^{3}$)
clusters \citep{cr05}. Only one of our clusters, M15, has such a high
density, and thus we do not a priori expect that such sources exist in our
other survey clusters.

The radius of the Arecibo beam at 1.4\,GHz is roughly $1.5^{\prime}$.  Thus,
for the 12 clusters we surveyed with $r_c < 0.5^{\prime}$, almost all of the
cluster's pulsars should have fallen within our beam.  Furthermore, the beam
radius of the 430-MHz \citet{and93} survey, which observed 11 of the same
sources (see Table~\ref{gcs.tab}), was $5^{\prime}$, and would likely have
detected some sources further from the cluster centers, if they existed. 
For the 6 clusters we surveyed with $r_c < 1.0^{\prime}$, the coverage was
still very good, though perhaps 20\% of the pulsar population fell
outside the Arecibo beam. There remain 4 survey clusters whose $r_c$ is
comparable to the beam half-power radius or larger: NGC~5053 ($r_c =
1.98^{\prime}$), NGC~5466 ($r_c = 1.64^{\prime}$), Pal~5 ($r_c =
3.25^{\prime}$), and Pal~15 ($r_c = 1.25^{\prime}$).  In these clusters,
perhaps only half of the pulsar population fell within the Arecibo beam. In
general however, we conclude that it is unlikely that a significant number
of pulsars were missed in this survey because of spatial coverage.

\subsection{MSP Spin Frequency Distribution}

%Based on a population of 22 field MSPs, Cordes and Chernoff 1997 find:
%Pmin > 1ms at 95% confidence and Pmin > 0.65ms @ 99% confidence
%dN/dP propto P^-2.0
%pseudoluminosity dist prop to L^-2.0+/-0.2
%luminosity cutoff L >= 1.1 mJy kpc^2 (@ 400MHz???)

%See Lorimer et al. 1996: 

%What is the intrinsic period dist?  47Tuc seems to have a deficit
%below 2ms.  Ter5 seems to have a deficit below 1.5ms.  For our survey,
%sensitivity to isolated pulsars starts to drop around 3\,ms, especially
%at higher DMs.

%Aspen Proc. Chakrabarty et al. : ``Recent radio pulsar surveys, in which 
%selection effects are accounted for, are independently finding similar
%evidence for a maximum spin frequency around 700 Hz, as reported at this
%eeting (McLaughlin et al., in this volume; Camilo \& Rasio, in this volume).''
%---> Maura finds a lower limit of 1.2ms from 11 pulsars found in AO drift
%surveys
%---> 

If one includes both the known MSPs in the field and those in GCs, the {\it
observed} distribution of radio pulsar spin frequencies above 200\,Hz
roughly follows a power-law relationship with an index of $-3$, i.e. $N_{\rm
psr} \propto {\nu_{\rm spin}}^{-3}$ (Figure~\ref{spinfdist.fig}, right). We do not
suggest that the underlying spin-frequency distribution is a power-law, or that
a physical motivation for this choice exists, we merely use this functional
form to quantify the sharp observed drop in the number of known pulsars as $\nu_{\rm
spin}$ increases.  Considering the combined spin frequencies of MSPs in the
field and in GCs is potentially problematic, as the frequency distributions
of MSPs in the field and those in GCs could well be intrinsically different.
Furthermore, the spin-frequency distribution between GC MSP populations may also
vary (e.g. there is some indication that Terzan~5 has a wider range of spin
frequencies than 47~Tucanae).  Nonetheless, for the purposes of discussing the
effect of observational bias on the observed spin frequency distribution, we
will consider the MSP population as a whole.

There is no significant correlation currently observed between the radio
luminosity of MSPs and their spin frequency.  Because all other conceivable
observational biases (e.g. scattering, dispersive smearing,
self-obscuration) {\it increase} the difficulty in detecting the fastest
pulsars, the observed spin-frequency distribution above 200\,Hz places a
limit on the steepness of the intrinsic spin frequency distribution at these
frequencies (i.e. if $N_{\rm psr} \propto {\nu_{\rm spin}}^{-\alpha}$, then
${\alpha}_{\rm true} < {\alpha}_{\rm obs}$). 
%\citet{cha05} \citep[see
%also][]{cmm+03} notes that the observed spin-frequency distribution of
%accreting LMXBs showing bursts (the ``nuclear-powered'' pulsars) is
%statistically consistent with a flat distribution between 270$-$619\,Hz, and a
%cut-off at 730\,Hz.  Given that LXMBs are the most likely progenitors of the
%MSPs, do MSPs also have a roughly flat spin-frequency distribution up to
%some cut-off frequency?
We now consider the observational biases that contribute to the
observed spin-frequency distribution of MSPs.  First, we see from
Equation~\ref{sens.eqn} that the minimum detectable flux density depends on
spin frequency because pulse smearing due to scattering and DM has a greater
relative effect for fast pulsars.  This accounts for part of the slope in
the observed spin-frequency distribution.  Second, as there is no indication
that orbital period is strongly correlated with spin frequency, it is unlikely
that many of the fastest MSPs are being missed because they are
preferentially in the most compact orbits. However, the number of bins
through which a binary-modulated pulsar signal will drift in the Fourier
domain is directly proportional to its spin frequency. In other words, for
the same orbital period, it is more difficult to detect a 1-ms pulsar than a
5-ms pulsar in an acceleration search, because the 1-ms pulsar (and its
harmonics) will drift by a factor of 5 times more Fourier bins.  Harmonics
that drift by many bins are more susceptible to non-linear frequency drift
terms (reducing their detectability) and may drift beyond the maximum number
of bins probed by the search ($z_{\rm max}$). Thus, binary motion also
accounts for part of the slope in the observed spin-frequency distribution,
although this effect is difficult to quantify.  Finally, because it is
plausible that eclipse fraction increases with $\dot{E} = 4 {\pi}^2 I \nu
\dot{\nu} \propto B_{\rm surf}^2 \nu^4$ (assuming that the pulsar spindown
is dominated by magnetic dipole radiation), fast-spinning pulsars in binary
systems may be preferentially obscured by the material their strong winds
ablate from their companions \citep{tav91,hrs+06}.

Given the multitude of observational biases against detecting the
fastest-spinning radio pulsars, it is difficult to identify what portion of
the observed spin-frequency distribution is instrinsic to the population.
Perhaps the best way to investigate this problem is by performing Monte
Carlo simulations of various trial underlying populations and then searching
these until one converges on the observed population. Terzan~5 and
47~Tucanae present excellent samples on which to perform such simulations,
although one would have to be careful in generalizing the results to MSPs in
the field or in other GCs.  If the distribution of radio pulsar spin
frequencies is even approximately like that of the LMXBs, which is roughly
consistent with being flat over the observed range of 270$-$619\,Hz
\citep{cmm+03}, then we have still only discovered a very
small fraction of the fastest-spinning radio pulsars. Mapping the
distribution at the fastest spin frequencies depends crucially on finding
these pulsars.

%Remove
%Radio surveys for rotation-powered pulsars will always be limited by
%scattering, but there is still room for higher time and frequency resolution
%backends (or coherent dedispersion) to improve sensitivity to the
%fastest-spinning pulsars. Surveys at higher observing frequencies than used
%here \cite[i.e. $> 1.5$\,GHz, e.g.][]{rhs+05b,rhs+05c} mitigate the effects
%of scattering, which scales as $\nu^{-4}_{\rm obs}$, and inter-channel
%dispersive smearing, which scales as $\nu^{-3}_{\rm obs}$.  Unfortunately,
%higher observing frequencies also require very large observing bandwidth to
%compensate for the typically steep spectral indices of pulsars.  X-ray
%surveys are also potentially interesting, as they do not suffer from the
%adverse effects of dispersion or scattering (although there is absorption of
%soft X-rays by intervening matter).  However, currently only one of the
%radio MSPs in GCs (PSR~B1821$-$24A) is also detected as an X-ray pulsar.
%Furthermore, the strong bias against detecting binaries can be mitigated by
%the development of increasingly sophisticated (and computationally
%intensive) search techniques that more completely correct for the pulsar's
%binary motion. Another advantage of future instruments, which will have
%greater instantaneous sensitivity, is that one may find faint pulsars in
%increasingly short data sets, where binary motion has less time to smear the
%signal.

\subsection{Pulsar Luminosities}

%Camilo \& Rasio: ``It is not straightforward to determine the luminosity
%function of the pulsars in 47 Tuc because of their large-amplitude
%scintillations''  From 14 pulsars, they find dlogN = -dlogL (e.g.
%McConnell et al. 2004).

%Camilo \& Rasio: ``Many surveys have a luminosity limit L1400 > 10 mJy kpc2,
%while the maximum luminosity for pulsars in 18 GCs is ~10mJy kpc2 (exceptions
%are the much brighter ..).  It seems therefore that in many cases at radio
%wavelengths that we are still only probing the tip of the iceberg.''

%Anderson maximum likelihood analysis is very sensitive to the Lmin
%that is used.

%Cordes & Chernoff 97: ``..we currently regard the similarity between
%long-period and MSP luminosity distributions as fortuitous''

%****Cordes & Chernoff 97: ``An analysis of GC MSP populations should
%probably use a treatment similar to this paper's but applied to cluster-only
%data.

%Cordes & Chernoff 97: ``On evolutionary grounds, many properties of disk
%and GC MSPs might be expected to differ (e.g. distributions of luminosity,
%spin period, orbital period, and velocity)''

%Cordes & Chernoff 97: ``Some observations suggest a weak positive correlation
%(Lundgren,Zepka & Cordes 1995) between orbital and spin period''

%Lyne et al. 1998 (Parkes Southern Pulsar Survey) find a dlogN=-dlogL
% relationship based on 21 MSPs, but have to make _very_ large corrections
% for observational bias and beaming.

In Table~\ref{pulsars.tab}, we list the 1.4-GHz flux densities of the
pulsars discovered in this survey, as well as those of the previously
known pulsars in these clusters.\footnote{We provide only upper limits,
determined from the raw sensitivity of the survey, on the flux
densities of the candidate pulsars M3C and NGC~6749B.}  These were
derived from the observed pulse profiles by integrating the pulse and
scaling this flux using the off-pulse root mean square and the
radiometer equation.  For the mostly non-scintillating pulsars M71A
and NGC~6749A, we estimate that the fractional uncertainty on their
flux is roughly 30\%.  For the pulsars found in M3, M5, and M13,
scintillation can have a strong effect on the observed flux of the
pulsars as a function of time, making it more difficult to estimate
the underlying intrinsic brightness of these sources.  For these
pulsars, we have used an approach similar to that used by
\citet{clf+00}: we fit the observed fluxes -- using half the survey
sensitivity limit as the flux in the case of non-detections -- to an
exponential distribution, whose median value we take as the intrinsic
flux.  This approach worked well for many of the scintillating pulsars
in M3, M5, and M13, which were detected in the majority of our many
($\gtrsim 50$) timing observations of these clusters.  For these
pulsars, we estimate that the fractional uncertainty on their quoted
flux density is 50\%.  The exceptions were M3A, M5E, and M13E, where
the scarcity of detections made determining the intrinsic flux density
more uncertain.  The quoted flux densities of these sources have a
higher fractional uncertainty of roughly 70\%.
%Furthermore, we derive the following 1400-MHz flux densities for the previously known
%pulsars in M5 and M13: \mva: 0.12\,mJy, \mvb: 0.025\,mJy, \mxiiia:
%0.14\,mJy, and \mxiiib: 0.022\,mJy.

The luminosity distribution of MSPs, both in the field and in GCs, has
been difficult to constrain precisely because of the relatively small
number of known MSPs and the observational biases against finding
faint, fast, and binary pulsars. Furthermore, individual estimates of
pulsar luminosity can suffer from large systematic errors because the
distance is incorrect.  This is especially difficult in the field,
where most distances are estimated from the DM, but is less of an
issue for pulsars associated with GCs, whose distances are known
comparatively precisely.  In this section, we take advantage of the
many new pulsar discoveries that have been recently made in GCs.  We
combine the luminosities of the pulsars currently known in M5, M13,
M15, M28, NGC~6440, NGC~6441, 47~Tucanae, and Terzan~5 (see
Table~\ref{lum_sample.tab}) and consider the resulting luminosity
distribution of GC MSPs.  These specific GCs were chosen because they
contain at least 4 pulsars each.  The clusters M3, M62, NGC~6624, and
NGC~6752, which are the only other clusters with at least 4 known
pulsars, were excluded from the analysis because reliable fluxes were
not available for all the known pulsars in these clusters.

\begin{figure}
\centerline{\includegraphics[height=1.0\columnwidth]{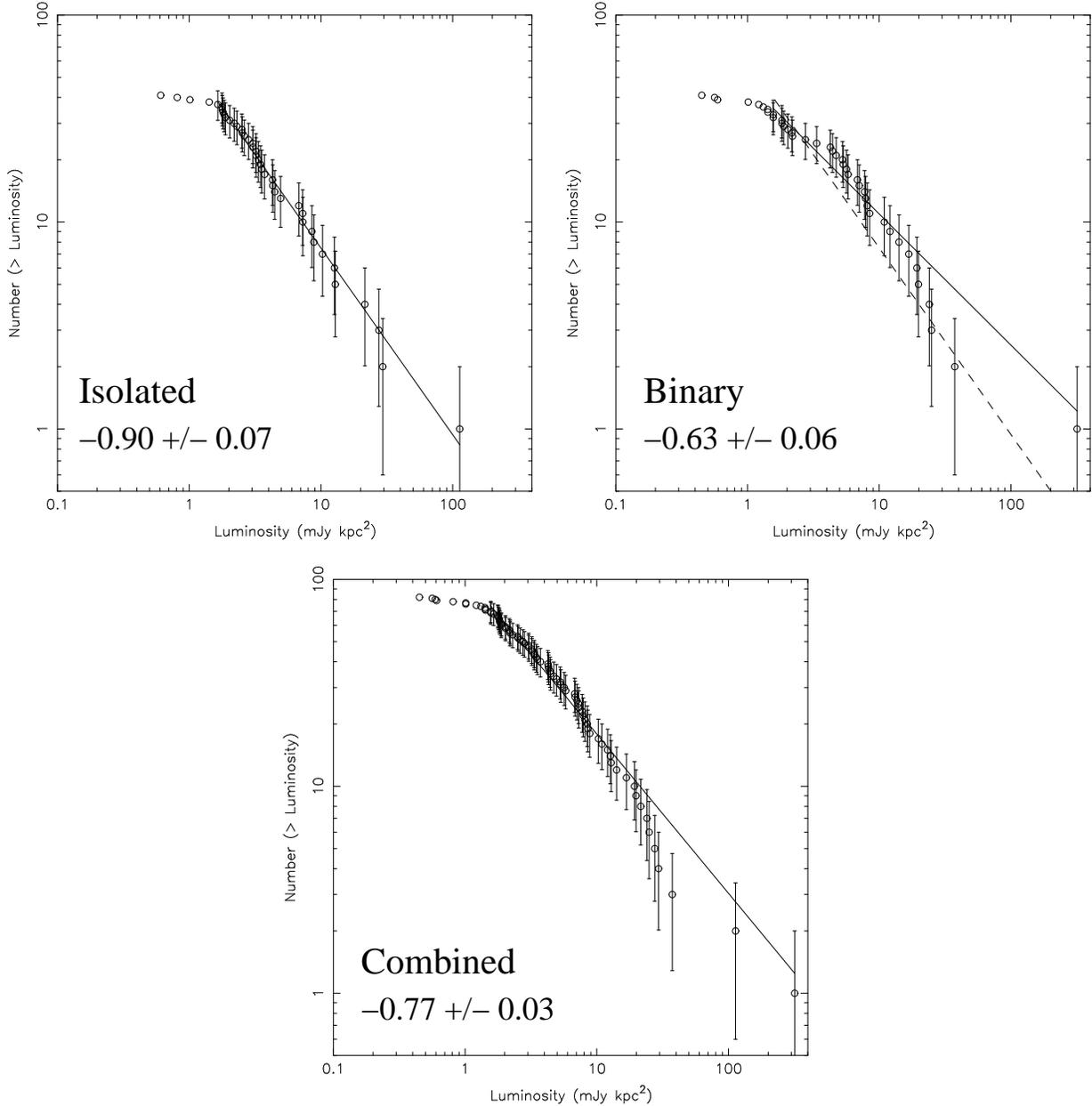}}
%\plotone{f5.eps}
\caption{Cumulative distributions of 1.4-GHz luminosities of pulsars in M5,
M13, M15, M28, NGC~6440, NGC~6441, 47~Tucanae, and Terzan~5.  The minimum
luminosity considered for fitting purposes is $L^{\rm min}_{1400} =
1.5$\,mJy kpc$^2$.  Error bars are the square-root of each value.  The
excluded points below $L^{\rm min}_{1400}$ are shown without error bars.
{\it Top left:} luminosity distribution of 41 isolated pulsars in these
clusters, which has a slope of $-0.90 \pm 0.07$ (solid line, 37 pulsars used
in fitting).  {\it Top right:} luminosity distribution of 41 binary pulsars
in these clusters, which has a slope of $-0.63 \pm 0.06$ (solid line, 33
pulsars used in fitting).  The best-fit slope from the distribution of
isolated pulsars is also shown overlaid as a dashed line.  {\it Bottom:}
combined luminosity distribution, including all 82 isolated and binary
pulsars.  The best-fit slope is $-0.77 \pm 0.03$ (solid line, 70 pulsars
used in fitting). \label{luminosities.fig}}
%McConnell 47 Tuc paper: -0.9 +/- 0.2 (Lmin_1400 ~ 1.4 mJy kpc2)
%Anderson thesis: -0.93 +/- 0.38 (Lmin_1400 ~ 1.5 mJy kpc2)
\end{figure}

In Figure~\ref{luminosities.fig}, we plot the 1.4-GHz cumulative
luminosity distribution of 41 isolated (top left), 41 binary (top
right), and all 82 pulsars in our sample combined (bottom).  In the
majority of cases, specific spectral indices were not available, and
luminosities were scaled to 1.4\,GHz (where necessary) assuming a
pulsar spectral index of $\alpha = -1.8$ \citep{mkkw00spec}.  Unless
otherwise indicated, the \citet{har96} catalog distance to the host GC
was used to convert 1400-MHz flux density $S_{1400}$ to pseudo
luminosity $L_{1400} \equiv S_{1400} d^{2}$ (see
Table~\ref{lum_sample.tab}). The M5 and M13 luminosities are from the
observations made in this survey. The Terzan~5 luminosities come from
\citet{rhs+05b} and subsequent analysis of the more recently found
pulsars in this cluster.  We assumed a distance of 8.7\,kpc to
Terzan~5 \citep{clge02}.  The 47~Tucanae fluxes are from
\citet{clf+00} and we use a distance of 4.5\,kpc \citep{zro+01} to
convert these to luminosities.  The luminosities of pulsars in M28,
NGC~6440, and NGC~6441 come from recent 1950-MHz discoveries and
timing observations made by our group with the GBT \citep[][these
discoveries are currently being prepared for publication as B\'egin et
al. and Freire et al.]{rhs+05c,beg06}.

We fit ${\rm log}_{10}(N > L)$ versus ${\rm log}_{10}(L)$ to a line, using
the square-root of ${\rm log}_{10}(N > L)$ as the uncertainties. These
best-fit slopes are shown as solid lines in Figure~\ref{luminosities.fig}.
No corrections have been made for any observational bias \citep[as has been
done for field MSPs in][]{lml+98,cc97}.  For all isolated and binary pulsars
combined, the best-fit slope is $-0.77 \pm 0.03$.  Below $1.5$\,mJy kpc$^2$
the observed distribution turns over, and thus we used a minimum luminosity
cut-off of $L^{\rm min}_{1400} = 1.5$\,mJy kpc$^2$ for fitting purposes.
Only the 70 pulsars in our sample above this luminosity limit were included
in the fit.  In combining luminosities from numerous clusters, we have
assumed that the luminosity function does not vary significantly between
clusters. We note that, when the same analysis is applied separately to the
pulsar populations of the individual GCs in our sample, the slope is in each
case consistent with that derived from all clusters in our sample combined
(though the error on the slope is of course large for clusters with few
known pulsars).  This supports the assumption that the radio luminosity
distribution of GC pulsars is universal. As 40\% of the sample pulsars are
in Terzan~5, it is important to ask how much the assumed distance to the
cluster affects the combined luminosity distribution. We find that the
distance to Terzan~5 can change by up to 30\%\footnote{This
encompasses the range of published distances to Terzan~5, including the
newest derived distance of $5.5 \pm 0.9$\,kpc from \citet{obb+07}.} without
significantly altering the slope derived from all the different cluster
pulsars combined.  Furthermore, 56\% of the pulsars in our sample are
contained in either Terzan~5 or 47~Tuc.  These two clusters thus have a
large influence on the derived luminosity law, and it will be important to
revisit these calculations when larger pulsar populations are known in other
clusters as well.

There are clearly ripples in the combined distribution, suggesting unmodeled
effects, possibly due entirely to observational biases.  When the population
is separated into isolated and binary pulsars, it is clear that these
effects come predominantly from the binary pulsars. This is perhaps
unsurprising, as the observational biases inherent in detecting such systems
are significantly higher than for the isolated pulsars.  Considering only
the 37 isolated pulsars in our sample above $L^{\rm min}_{1400}$, we find a
much smoother distribution, with a best-fit slope $-0.90 \pm 0.07$.
Conversely, the cumulative distribution of the 33 binaries above $L^{\rm
min}_{1400}$ is relatively poorly fit by a single slope of $-0.63 \pm 0.06$.
We note that while the slope derived by fitting the isolated pulsars is
relativity insensitive to the value of $L^{\rm min}_{1400}$ (as long as it
is not well below $1.5$\,mJy kpc$^2$), the slope derived from the binary
pulsars varies significantly as $L^{\rm min}_{1400}$ is changed. 

%If the luminosities of the
%binary pulsars also follow a single power law, then it appears that there
%are roughly three pulsars with $L_{1400} \grtsim 20$\,mJy kpc$^2$ missing
%from our sample.  Such pulsar should be easier to discover because of their
%brightness, unless they are preferentially subject to self-obscuration by a
%bloated companion (e.g. Terzan5~A).  Another possibility is that the
%brightest pulsars also have the highest spin-down luminosity and hence
%ablate their companion more rapidly and become isolated.  This is
%speculative however, as it has not been shown that radio luminosity scales
%with spin-down luminosity and the rate of companion ablation also depends on
%other factors, such as orbital separation. 

Given the much lower bias against detecting isolated pulsars, we suggest
that the slope derived from fitting only isolated pulsars is the most
reliable. This distribution is somewhat flatter, though still barely
consistent with the d ${\rm log}_{10}(N)$ = $-$ d ${\rm log}_{10}(L)$
relation found for non-recycled field pulsars and MSPs \citep*[consider for
instance][]{lmt85,lml+98,cc97}. It is also roughly consistent with the
recently derived luminosity law of \citet{lfl+06}, who find d ${\rm
log}_{10}(N) \sim -0.8$ d ${\rm log}_{10}(L)$ using a sample of 1008 normal
(non-millisecond) pulsars. As the population of known GC MSPs continues to
increase, we will be better able to constrain the luminosity distribution. 
However, until the next large advance in telescope collecting area, it will
be hard to constrain the luminosity function of GC MSPs below 1\,mJy kpc$^2$,
because of the relatively large distances to GCs and selection effects,
which are worse for weak pulsars.

We note that 70\% of our sample pulsars, those in M15, M28, NGC~6440,
NGC~6441, and Terzan~5, can be characterized as not strongly scintillating
(Table~\ref{lum_sample.tab}).  Thus, as the majority of the pulsars in the
sample don't scintillate significantly, and those that do have been given
special attention, we don't believe that scintillation is strongly biasing our
distribution.  When we consider only the 31 not strongly scintillating,
isolated pulsars above our luminosity cutoff, we find a best-fit slope of
$-0.86 \pm 0.08$, consistent with that derived from all our isolated sample
pulsars.  We note further that since the known MSPs in the plane have mostly
very low DMs ($\sim 80$\% have DM$ < 50$\,pc cm$^{-3}$) and are observed for
shorter amounts of time (which means less averaging over strong
scintillation epochs), scintillation is potentially a larger pitfall in
analyzing the luminosities of those sources.

As the flux densities used in this analysis were obtained at either
0.43\,GHz, 1.4\,GHz, or 2.0\,GHz (see Table~\ref{lum_sample.tab}) we have
also investigated the uncertainty introduced by the error on the mean
spectral index used to scale these fluxes.  The nominal error on the $-1.8$
mean spectral index we have used is 0.2 \citep{mkkw00spec}. We have re-run
our fitting of isolated pulsars using a spectral index of -1.6 and -2.0 and
find slopes of $-0.92 \pm 0.07$ and $-0.89 \pm 0.07$ respectively for
isolated pulsars.  Thus, uncertainty in the average spectral index of these
pulsars does not have a large effect on the derived luminosity distribution.

%Notes on Bailes et al. 1997
%
%KS test shows that the isolated and binary luminosity dists are different
%at the 99.5% level.
%Distance estimates are from TC93
%Lum at 400MHz
%They say that uncertainty in distance makes certain quantities hard
%to compare between iso and bin MSPs{\ldots}  why not also lums, which
%depend on d^2?
%''Either there is a selection effect or the int lum of iso MSPs are
%lower on average than bins
%''We are therefore left with the possibility that some fraction of the
%solitary MSPs form by a completely diff mech from that of bin MSPs

%Note on Lommen et al. 2006
%
%Find a diff in scale height between iso and bin MSPs
%Find no diff in vels
%Scale height diff due to diff in lums?
%Most distances from NE2001 (CL01)
%Confirm the Bailes et al. result with an updated catalog
%Also, the median dist of iso is 510pc, while that of bin is 1155pc,
%also suggests that the iso MSPs must be less luminous.
%Lum diff between iso and bin also suggested by
%Kramer et al. 1998, Hobbs et al. 2004

Lastly, we have statistically compared the luminosity distributions of
isolated and binary GC pulsars, using a Kolmogorov-Smirnov (KS) test
\citep[e.g.][]{pft86}.  We find that the two populations are statistically
different with only 34\% confidence. In other words, there is currently no
evidence that the observed luminosity distributions of isolated and binary
GC pulsars are significantly different.  This can also be seen by plotting
the best-fit slope from the distribution of isolated GC pulsars over the
distribution of binaries (dashed-line in Figure~\ref{luminosities.fig}, top
right).  This lack of difference has also been recently demonstrated for the
MSPs in the Galactic plane by \citet{lmcs07}, who argue that previous claims
\citep{bjb+97spec,lkn+06} of a statistical difference between the luminosities
of the two populations were due to small number statistics and observational
biases.

%Superceded
%This
%lack of obvious difference contrasts with what has been found for isolated
%and binary MSPs in the Galactic plane, where \citet{bjb+97spec} show, also on
%the basis of a KS test, that the observed luminosity functions of the two
%populations differ with 99.5\% confidence.\footnote{This result has been
%confirmed with a more recent catalog of field MSPs by \citet{lkn+06}.}
%The lack of observed difference in the luminosity distributions of isolated
%and binary pulsars in GCs may be due to collisions and exchange interactions
%in the dense core of the cluster, which could wash out luminosity
%differences due to purely the formation of the original MSP.  Alternately,
%it may be possible that the observed difference between the luminosities of
%isolated and binary MSPs in the field is due to selection effects against
%finding dim binaries.

\section{Conclusions}

We have used the Arecibo radio telescope at $\sim 1.4$\,GHz to search 22 GCs
for pulsars.  These searches are among the deepest searches ever undertaken
for such objects, and employed the most sensitive algorithms available to
find Doppler-shifted binary pulsar signals.  Our survey discovered 11 MSPs,
almost doubling the known population in these GCs.  8 of these new pulsars
are in binary systems, and 3 show eclipses. We find a significantly higher
proportion of binaries, eclipsing systems, and pulsars with very short spin
periods ($P_{\rm spin} < 4$\,ms) than previous searches of these clusters.
We consider the luminosity distribution of GC pulsars and find that these
follow a form d ${\rm log}_{10}(N)$ = $-0.90 \pm 0.07$ d ${\rm
log}_{10}(L)$.  We find no evidence for a difference in the luminosity
distributions of isolated and binary GC pulsars.

\acknowledgements

J.W.T.H. thanks NSERC for a PGS-D fellowship, which was tenured during this
research.  I.H.S. holds an NSERC UFA and is supported by an NSERC Discovery
Grant. V.M.K. is a Canada Research Chair, and acknowledges support from an
NSERC Discovery Grant and Steacie Supplement, CIAR, and from the FQRNT. We
sincerely acknowledge Arun Venkataraman, Jeff Hagen, and Bill Sisk of the
Arecibo Observatory for their fantastic help with data management and
maintenance of the WAPPs.  We also thank Dunc Lorimer for his valuable
guidance in our early days of WAPP use and Fernando Camilo, our referee, who
provided detailed comments and suggestions which improved our original
manuscript.  V.M.K., J.W.T.H., and S.M.R. are very grateful to the Canada
Foundation for Innovation for the New Opportunities Grant that funded
construction of ``The Borg'', the computer cluster that was essential for
our analysis, and to Paul Mercure for helping to maintain this system. The
Arecibo Observatory is part of the National Astronomy and Ionosphere Center,
which is operated by Cornell University under a cooperative agreement with
the National Science Foundation. 

%%%%%%%%%%%%%%%%
% Bibliography %
%%%%%%%%%%%%%%%%

%\bibliographystyle{apj}
%\bibliography{journals1,modrefs,psrrefs,AOGC_refs}

%%%%%%%%%%
% TABLES %
%%%%%%%%%%

%TABLE: Survey Source List and Observations
\newpage

\begin{deluxetable}{llcccccccccc}
\rotate
%\tabletypesize{\footnotesize}
\tabletypesize{\scriptsize}
\tablecolumns{8}
\tablewidth{0pc}
\tablecaption{Survey Source List and Observations\label{gcs.tab}}
\tablehead{
%Header first line
\colhead{Catalog ID} 
& \colhead{Other} 
& \colhead{Distance}
& \colhead{$r_c$\tablenotemark{a}}
& \colhead{$\rho_{\circ}$\tablenotemark{b}}
& \colhead{Concentration\tablenotemark{c}}
& \colhead{DM($\Delta$DM)\tablenotemark{d}} 
& \colhead{Predicted DM\tablenotemark{d}} 
& \colhead{Time Visible\tablenotemark{e}}
& \colhead{Isolated\tablenotemark{f}} 
& \colhead{Binary\tablenotemark{f}}  
& \colhead{Lum. Limit\tablenotemark{g}} 
\\
%Header second line
& \colhead{Name}
& \colhead{(kpc)}   
& \colhead{(arcmin)}   
& \colhead{(${\rm log}_{10}$ \lsun/pc$^{3}$)}
& \colhead{(${\rm log}_{10}$ $r_t/r_c$)}
& \colhead{(pc cm$^{-3}$)} 
& \colhead{(pc cm$^{-3}$)} 
& \colhead{(hrs)}
& \colhead{Pulsars} 
& \colhead{Pulsars}
& \colhead{(mJy kpc$^2$)}
}
\startdata

NGC~4147$^*$ & \nodata & 19   & 0.10 & 3.48    & 1.80 & \nodata    &  24  & 2.8 & \nodata & \nodata & 6.8\\
NGC~5024$^*$ & M53     & 18   & 0.36 & 3.05    & 1.78 & 24         &  25  & 2.8 & \nodata & 1       & 6.1\\
NGC~5053     & \nodata & 16   & 1.98 & 0.53    & 0.84 & \nodata    &  25  & 2.7 & \nodata & \nodata & 5.0\\
NGC~5272$^*$ & M3      & 10   & 0.55 & 3.51    & 1.84 & 26.4(0.3)  &  23  & 2.5 & \nodata & 4(4)    & 2.1\\
NGC~5466     & \nodata & 16   & 1.64 & 0.88    & 1.32 & \nodata    &  22  & 2.4 & \nodata & \nodata & 5.7\\
NGC~5904$^*$ & M5      & 7.5  & 0.42 & 3.91    & 1.83 & 29.5(0.8)  &  32  & 1.5 & 1 & 4(3)    & 1.4\\
NGC~6205$^*$ & M13     & 7.7  & 0.78 & 3.33    & 1.51 & 30.2(1.1)  &  38  & 1.2 & 2(1)  & 3(2)    & 1.7\\
NGC~6426$^*$ & \nodata & 21   & 0.26 & 2.35    & 1.70 & \nodata    & 121  & 1.7 & \nodata & \nodata & 11\\
NGC~6535$^*$ & \nodata & 6.8  & 0.42 & 2.69    & 1.30 & \nodata    & 172  & 0.8 & \nodata & \nodata & 1.9\\
NGC~6749     & Be42    & 7.9  & 0.77 & 3.33    & 0.83 & 193(2)     & 438  & 1.5 & \nodata & 2(2)    & 2.0\\
NGC~6760$^*$ & \nodata & 7.4  & 0.33 & 3.84    & 1.59 & 200(6)     & 257  & 1.3 & 1       & 1       & 2.0\\
NGC~6779     & M56     & 10   & 0.37 & 3.26    & 1.37 & \nodata    & 163  & 2.3 & \nodata & \nodata & 2.5\\
NGC~6838     & M71     & 4.0  & 0.63 & 3.04    & 1.15 & 117        & 86   & 2.8 & \nodata & 1(1)  & 0.3\\
NGC~6934$^*$ & \nodata & 16   & 0.25 & 3.43    & 1.53 & \nodata    & 83   & 2.2 & \nodata & \nodata & 6.6\\
NGC~7006$^*$ & \nodata & 42   & 0.24 & 2.46    & 1.42 & \nodata    & 74   & 2.7 & \nodata & \nodata & 34\\
NGC~7078$^*$ & M15     & 10   & 0.07 & 5.38    & 2.50 & 66.9(2.2)  & 68   & 2.6 &  7 & 1       & 2.1\\
NGC~7089     & M2      & 12   & 0.34 & 3.90    & 1.80 & \nodata    & 46   & 0.6 & \nodata & \nodata & 5.3\\
Pal~2        & \nodata & 28   & 0.24 & 3.76    & 1.45 & \nodata    & 136  & 2.3 & \nodata & \nodata & 18\\
Pal~5        & \nodata & 23   & 3.25 & $-$0.81 & 0.70 & \nodata    & 34   & 0.9 & \nodata & \nodata & 17\\
Pal~10       & \nodata & 5.9  & 0.81 & 3.50    & 0.58 & \nodata    & 166  & 2.8 & \nodata & \nodata & 0.8\\
Pal~13       & \nodata & 26   & 0.65 & 0.40    & 0.68 & \nodata    & 38   & 2.6 & \nodata & \nodata & 14\\
Pal~15       & \nodata & 45   & 1.25 & $-$0.27 & 0.60 & \nodata    & 77   & 0.7 & \nodata & \nodata & 76\\
\enddata
\vspace{-0.5cm}
\tablecomments{Clusters marked with an asterisk (*) were also searched by
\citet{and93} with Arecibo at 430\,MHz.  \citet{and93} also searched
NGC~6218 (M12, no pulsars found), which is not visible using the Arecibo
Gregorian dome.  All previously known pulsars were found by \citet{and93},
with the exceptions of PSR~J1911+0101B \citep[][in NGC~6760]{fhn+05},
PSRs~B1516+02A and B \citep[][in M5]{wak+89a}, and PSR~B2127+11A \citep[][in
M15]{wkm+89b}.} 
\tablenotetext{a}{The core radius of the cluster.  For comparison, the
half-power radius of the Arecibo 1.4-GHz beam is $\sim 1.5^{\prime}$.}
\tablenotetext{b}{The logarithm of the central luminosity density of the cluster.}
\tablenotetext{c}{The logarithm of the ratio of the tidal radius to the core radius of the
cluster.} 
\tablenotetext{d}{Predicted
values of DM are based on the \citet{cl02spec} electron density model of the Galaxy.
There is no formal uncertainty on these values, and the predicted DM
can sometimes differ from the true value by a factor of two or more.  The
predicted DM is also provided for clusters with known pulsars, and this
shows the characteristic discrepancy between the values.  In some cases, the
predicted DM agrees very well with the DM of pulsars in the cluster.  This
is because these pulsars have been used to scale the model itself.  For
clusters with more than one known pulsar, $\Delta {\rm DM}$ indicates the
observed spread in DM.}
\tablenotetext{e}{Time visible with the Arecibo telescope, which can only
track sources while they are within $20^{\circ}$ of the zenith.}
\tablenotetext{f}{Values in parentheses indicate the number of pulsars in
the cluster that were found in this survey.  The numbers for NGC~5272 (M3)
and NGC~6749 include one cadidate pulsar (yet to be confirmed) each.}
\tablenotetext{g}{Approximate upper limit on the pseudo-luminosity at
1400\,MHz ($L_{1400} = S_{1400} d^{2}$) of an undiscovered pulsar in the
cluster.  This assumes that the pulsar has a spin period $\gtrsim 1$\,ms and
is isolated.}
\end{deluxetable}

%TABLE: Comparison of observing systems at Arecibo, Parkes, and GBT
\newpage

\begin{deluxetable}{lcccc}
\tablecolumns{5}
\tablewidth{0pc}
\tablecaption{Comparison of Recent GC Pulsar Surveys\label{surveys_comp.tab}}
\tablehead{
%Header first line
\colhead{Telescope} 
& \colhead{Gain} 
& \colhead{Bandwidth} 
& \colhead{Int. Time} 
& \colhead{Obs. Freq.} 
\\
%Header second line
& \colhead{(K/Jy)} 
& \colhead{(MHz)} 
& \colhead{(hr)} 
& \colhead{(GHz)} 
}
\startdata
%Using \phantom to align decimals
Arecibo & 10.5 & 100 & 2 & 1.4 \\
Parkes  & \phantom{1}0.7 & 288 & 8 & 1.4 \\
GBT     & \phantom{1}2.0 & 600 & 8 & 2.0 \\
\enddata
\tablecomments{Arecibo parameters are for the survey described here.  Parkes
parameters are for the Multibeam Filterbank system, used in a number of recent 
GC searches there \citep[e.g.][]{clf+00}.  GBT parameters are for surveys using 
the Pulsar Spigot \citep[e.g.][]{rhs+05b}.}
\end{deluxetable}

%TABLE: Previously Known Pulsars
\newpage

\begin{deluxetable}{llcccccccc}
\rotate
\tablecolumns{6}
\tablewidth{0pc}
\tablecaption{Previously Known Pulsars\label{prev_pulsars.tab}}
\tablehead{
%Header first line
\colhead{Name} 
& \colhead{Period} 
& \colhead{Isolated (I) or} 
& \colhead{Redetected?} 
\\
%Header second line
\colhead{Informal / Formal} 
& \colhead{(ms)} 
& \colhead{Binary (B)} 
& \colhead{(Y/N)} 
}
\startdata
%Using \phantom to align decimals
M5A / \mva        & \phantom{11}5.554 & I & Y   \\
M5B / \mvb        & \phantom{11}7.947 & B & Y   \\
M13A / \mxiiia    & \phantom{1}10.378 & I & Y   \\
M13B / \mxiiib    & \phantom{11}3.528 & B & Y   \\
M15A / \mxva      & 110.665 & I & Y   \\
M15B / \mxvb      & \phantom{1}56.133 & I & Y   \\
M15C / \mxvc      & \phantom{1}30.529 & B & Y   \\
M15D / \mxvd      & \phantom{11}4.803 & I & Y   \\
M15E / \mxve      & \phantom{11}4.651 & I & Y   \\
M15F / \mxvf      & \phantom{11}4.027 & I & N\tablenotemark{a} \\
M15G / \mxvg      & \phantom{1}37.660 & I & N   \\
M15H / \mxvh      & \phantom{11}6.743 & I & N   \\
M53A / \mliiia    & \phantom{1}33.163 & B & Y   \\
NGC~6760A / \nsixa & \phantom{11}3.619 & B & Y   \\
NGC~6760B / \nsixb & \phantom{11}5.384 & I & Y   \\
\enddata
\tablenotetext{a}{Detected in 327-MHz WAPP data, see \S4.2.2.}
\end{deluxetable}

%TABLE: Pulsars and Their Basic Parameters
\newpage

\begin{deluxetable}{lccccccccc}
\rotate
\tabletypesize{\scriptsize}
\tablecolumns{10}
\tablewidth{0pc}
\tablecaption{Pulsars and Their Basic Parameters\label{pulsars.tab}}
\tablehead{
%Header first line
\colhead{Name\tablenotemark{a}} 
& \colhead{Period} 
& \colhead{DM} 
& \colhead{$P_{\rm orbit}$} 
& \colhead{$a_1 \sin (i)/c$}
& \colhead{Min $M_2$\tablenotemark{b}}
& \colhead{$w_{50}$\tablenotemark{c}}
& \colhead{Flux Density\tablenotemark{d}}
& \colhead{Number}
& \colhead{Span}
\\
%Header second line
\colhead{Informal / Formal}
& \colhead{(ms)} 
& \colhead{(pc cm$^{-3}$)} 
& \colhead{(hr)} 
& \colhead{(lt-s)} 
& \colhead{(\msun)}
& \colhead{(\%)}
& \colhead{($\mu$Jy)}
& \colhead{Det./Obs.}
& \colhead{(MJD)}
}
\startdata
M3A / \miiia                          & \phantom{1}2.545 & 26.5 & Unk. & Unk. & Unk. & 9.3 & 7 & 3/78 & 52491$-$52770\\
M3B / \miiib                          & \phantom{1}2.389 & 26.2 & 34.0 & 1.88  & 0.21 & 8.2 & 14 & 16/78 & 52485$-$53335\\
M3C\tablenotemark{e} / \miiic         & \phantom{1}2.166 & 26.5 & Unk. & Unk. & Unk. & 11 & $\lesssim 6$ & 1/78 & 52337\\
M3D / \miiid                          & \phantom{1}5.443 & 26.3 & 129\,d & 38.5 & 0.21 & 9.2 & 10 & 12/78 & 52768$-$53149\\
{\it M5A\tablenotemark{f} / \mva}     & \phantom{1}5.554 & 30.1 & \nodata & \nodata & \nodata & 6.7 & 120 \\
{\it M5B / \mvb}                      & \phantom{1}7.947 & 29.5 & 165 & 3.05  & 0.11 & 20 & 25 \\
M5C\tablenotemark{g} / \mvc           & \phantom{1}2.484 & 29.3 & 2.08 & 0.0573 & 0.038 & 6.2 & 39 & 59/60 & 52087$-$53422\\
M5D / \mvd                            & \phantom{1}2.988 & 29.3 & 29.3 & 1.60 & 0.20 & 18 & 8 & 24/60 & 52090$-$53335\\
M5E / \mve                            & \phantom{1}3.182 & 29.3 & 26.3 & 1.15 & 0.15 & 7.9 & 10 & 12/60 & 52705$-$53399\\
{\it M13A\tablenotemark{f} / \mxiiia} & 10.378 & 30.4 & \nodata & \nodata & \nodata & 11 & 140 \\
{\it M13B / \mxiiib}                  & \phantom{1}3.528 & 29.5 & 30.2 & 1.39 & 0.16 & 13 & 22 \\
M13C\tablenotemark{f} / \mxiiic       & \phantom{1}3.722 & 30.1 & \nodata & \nodata & \nodata & 5.5 & 30 & 60/67 & 52087$-$53422\\
M13D / \mxiiid                        & \phantom{1}3.118 & 30.6 & 14.2 & 0.924 & 0.18 & 6.6 & 24 & 63/67 & 52087$-$53422\\
M13E\tablenotemark{g} / \mxiiie       & \phantom{1}2.487 & 30.3 & $2.8 \pm 0.2 $ & $0.037 \pm 0.004$ & 0.02 & 6.0 & 10 & 2/67 & 52892,52833\\
M71A\tablenotemark{g} / \mlxxia       & \phantom{1}4.888 & 117  & 4.24 & 0.0782 & 0.032 & 10 & 59 & 53/53 & 52082$-$53422\\
NGC~6749A / \nvia                      & \phantom{1}3.193 & 194  & 19.5 & 0.588 & 0.090 & 16 & 23 & 17/17 & 52494$-$53 \\
NGC~6749B\tablenotemark{e} / \nvib     & \phantom{1}4.968 & 192  & Unk. & Unk.  & Unk. & 7.8 & $\lesssim 6$ & 1/17 & 52921\\
\enddata
\tablecomments{Unless otherwise indicated, errors on quantities are
well below the level of the least significant figure quoted.}
\tablenotetext{a}{Italicized names are those of the previously known pulsars in clusters
where new pulsars have been found.}
\tablenotetext{b}{Assuming a pulsar mass ($M_1$) of 1.4\,\msun\ and orbital inclination $i = 90^{\circ}$.}
\tablenotetext{c}{For profiles with multiple components, the width of the
highest peak is given.  As all the profiles suffer from residual dispersive
smearing, these values represent upper limits on the intrinsic pulse width at this
observing frequency.}
\tablenotetext{d}{Flux density at 1400\,MHz.  The approximate fractional uncertainty 
ranges from $30-70$\% depending on whether the source stronly scintillates and, if so, 
how often it was detected (see \S5.3).}
\tablenotetext{e}{Unconfirmed candidate pulsar.}
\tablenotetext{f}{Isolated.}
\tablenotetext{g}{Eclipses.}
\end{deluxetable}

%M3B & NGC6749A Timing Solutiona
\newpage

\begin{deluxetable}{lccc}
\tabletypesize{\footnotesize}
\tablecolumns{4}
\tablewidth{0pc}
\tablecaption{Timing Solutions\tablenotemark{*}~~for M3B, M3D, and NGC~6749A \label{M3pulsars.tab}}
\tablehead{ \colhead{} & \colhead{\miiib} & \colhead{\miiid} & \colhead{\nvia} }
\startdata
\multicolumn{4}{c}{Observation and data reduction parameters}\\
\hline
Period Epoch (MJD)   & 52770 & 52770 & 52770 \\
Start time (MJD)     & 52763 & 52768 & 53070 \\
End time (MJD)       & 53542 & 53476 & 54210 \\
\# of TOAs  & 161 & 83 & 77 \\
TOA rms ($\mu$s) & 9.1 & 24 & 38 \\
\hline
\multicolumn{4}{c}{Timing parameters}\\
\hline
$\alpha$\tablenotemark{a}& $13^{\rm h}42^{\rm m}11\fs 0871(1)$ & $13^{\rm
h}42^{\rm m}10\fs 2(6)$ & $19^{\rm h}05^{\rm m}15\fs 4(4)$ \\
$\delta$ & $+28^\circ 22\arcmin 40\farcs 141(2)$ & $+28^\circ 22\arcmin
36(14)^{\prime \prime}$ & $+01^\circ 54\arcmin 33(22)^{\prime \prime}$ \\
$P$ (ms)        & 2.389420757786(1) & 5.44297516(6) & 3.19294082(1) \\
$\dot{P}_{\rm obs}$ (10$^{-20}$) & 1.858(4) & $-$ & $-$ \\
DM ($\rm cm^{-3}\,pc$) & 26.148(2) & 26.34(2) & 193.692(8) \\
$P_b$(days)\tablenotemark{b} & 1.417352298(2) & 128.752(5) & 0.81255243(2) \\
$T_{\rm asc}$ (MJD) & 52485.9679712(6) & 52655.38(4) & 52493.83300(4) \\
$x$ (s)         & 1.875655(2) & 38.524(4) & 0.58862(2) \\
$e$             & $-$ & 0.0753(5) & $-$ \\
\hline
\multicolumn{4}{c}{Derived parameters}\\
\hline
$\theta_{\perp}$ (\arcmin) & 0.14 & 0.23(12) & 0.51(38) \\
%$a_{z \rm C}$ (m\,s$^{-2}$) & $<\,-2.6\,\times\,10^{-10}$ & $<+\,1.6\,\times\,10^{-10}$ \\
$\dot{P}_{\rm int}$ & $<\,3.4\,\times\,10^{-20}$ & $-$ & $-$ \\
$\tau_c$ (Gyr) & $>\,1.1$ & $-$ & $-$ \\
$B_0$ (gauss) & $<\,8.2\,\times\,10^7$ & $-$ & $-$ \\
$f\,(\rm M_{\odot})$ & 0.003526842(6) & 0.0037031(8) & 0.00033132(7) \\
$m_c$$(\rm M_{\odot})$\tablenotemark{c} & 0.21 & 0.21 & 0.090 \\
\enddata
\tablenotetext{*}{Note that the solutions presented for M3D and NGC~6749A are {\it not}
completely phase-connected solutions.  Due to sparse sampling, arbitrary phase jumps were used
between some observing epochs (see \S4.2.1).}
\tablenotetext{a}{The uncertainties indicated for all parameters
are twice the formal values given by {\tt TEMPO}.
We have used the Jet Propulsion Laboratory's DE405 planetary ephemeris
\citep{stan98} to derive these solutions.}
\tablenotetext{b}{The orbital parameters are: orbital
period ($P_b$), time of passage through the ascending node ($T_{\rm
  asc}$), semi-major axis of the orbit of the pulsar, projected along
the line-of-sight, divided by the speed of light ($x$) and orbital
eccentricity ($e$). Since the latter quantity is too small to be
measured, we can not estimate the longitude of the periastron relative
to ascending node ($\omega$). All other parameters are as described in
the text.}
\tablenotetext{c}{To calculate the minimum companion mass $m_c$, we assumed an
inclination angle $i=90^\circ$ and a pulsar mass of
1.4~M$_{\odot}$.}
\end{deluxetable}

%TABLE: Sample of pulsars used for luminosity distribution
\newpage

\begin{deluxetable}{lccccc}
\tablecolumns{6}
\tablewidth{0pc}
\tablecaption{Sample of Pulsars Used for Luminosity Distribution\label{lum_sample.tab}}
\tablehead{
%Header first line
\colhead{Cluster} 
& \colhead{Distance} 
& \colhead{DM} 
& \colhead{\# Pulsars} 
& \colhead{Obs. Freq.}
& \colhead{Scintillating?} 
\\
%Header second line
& \colhead{(kpc)} 
& \colhead{(pc cm$^{-3}$)} 
& \colhead{} 
& \colhead{(GHz)}
& \colhead{(Y/N)} 
}
\startdata
M5       & 7.5  & 29.5  & 5  & 1.4 & Y \\
M13      & 7.7  & 30.2  & 5  & 1.4 & Y \\
M15      & 10.3 & 66.9  & 8  & 0.4 & N \\
M28      & 5.6  & 120.5 & 8  & 2.0 & N \\
NGC~6440 & 8.4  & 223.4 & 5  & 2.0 & N \\  
NGC~6441 & 11.7 & 231.8 & 4  & 2.0 & N \\
Terzan~5 & 8.7  & 238.0 & 33 & 1.4 & N \\
47~Tuc   & 4.5  & 24.3  & 22 & 1.4 & Y \\
\enddata
\end{deluxetable}
  
%%%%%%%%%%%
% FIGURES %
%%%%%%%%%%%

%Survey Sensitivity
\newpage

%Sample Candidate Plot
\newpage

%Pulse Profiles
\newpage

%Spin Frequency Distributions
\newpage

%Luminosity Distributions
\newpage

\end{document}